\documentclass[%
 reprint,
 superscriptaddress,
amsmath,
amssymb,
aps,
pra,
]{revtex4-2}

\usepackage{graphicx}
\usepackage{nicefrac}
\usepackage{dcolumn}
\usepackage{bm}
\usepackage[hidelinks]{hyperref}
\usepackage{multirow}
\usepackage[raggedright]{subfigure}

\usepackage{physics}
\usepackage{bm}
\usepackage{pifont}
\newcommand{\cmark}{\ding{51}} 
\newcommand{\xmark}{\ding{55}} 
\usepackage{tikz}
\usepackage{blindtext}
\usepackage[acronym]{glossaries}
\usepackage{physics}
\usepackage[capitalize,noabbrev]{cleveref}
\Crefname{equation}{Eq.}{Eqs.}
\Crefname{figure}{Fig.}{Figs.}
\Crefname{section}{Sec.}{Secs.}
\Crefname{table}{Tab.}{Tabs.}
\crefname{appsec}{App.}{Apps.}

\begin{document}

\newacronym{vqc}{VQC}{variational quantum circuit}
\newacronym{cptp}{CPTP}{completely positive trace preserving}
\newacronym{qec}{QEC}{quantum error correction}
\newacronym{varqec}{VarQEC}{variational quantum error correction}
\newacronym{rea}{REA}{randomized entangling ansatz}
\newacronym{ml}{ML}{machine learning}
\newacronym{ftqc}{FTQC}{fault-tolerant quantum computing}
\newacronym{eftqc}{EFTQC}{early fault-tolerant quantum computing}
\newacronym{vareftqc}{VarEFTQC}{variational early fault-tolerant quantum computing}
\newacronym{iqp}{IQP}{instantaneous quantum polynomial}
\newacronym{mub}{MUB}{mutually unbiased bases}
\newacronym{sicpovm}{SIC-POVM}{symmetric informationally complete positive operator-valued measure}

\preprint{APS/123-QED}

\title{Learning Logical Operations for Arbitrary Quantum Error Correction Codes} 

\author{Nico Meyer}
\email{nico.meyer@iis.fraunhofer.de}
\affiliation{Fraunhofer IIS, Fraunhofer Institute for Integrated Circuits IIS, Nuremberg, Germany}
\affiliation{Pattern Recognition Lab, Friedrich-Alexander-University Erlangen-Nuremberg, Erlangen, Germany}

\author{Christopher Mutschler}
\affiliation{Fraunhofer IIS, Fraunhofer Institute for Integrated Circuits IIS, Nuremberg, Germany}
\affiliation{University of Technology Nuremberg (UTN), Nuremberg, Germany}

\author{Dominik Seuß}
\affiliation{Fraunhofer IIS, Fraunhofer Institute for Integrated Circuits IIS, Nuremberg, Germany}
\affiliation{Center for Artificial Intelligence (CAIRO), Technical University of Applied Sciences Würzburg-Schweinfurt, Würzburg, Germany}

\author{Andreas Maier}
\affiliation{Pattern Recognition Lab, Friedrich-Alexander-University Erlangen-Nuremberg, Erlangen, Germany}

\author{Daniel D. Scherer}
\affiliation{Fraunhofer IIS, Fraunhofer Institute for Integrated Circuits IIS, Nuremberg, Germany}

\begin{abstract}
\noindent
Logical operations are essential for quantum computation within quantum error-correcting codes. However, discovering their physical realizations is challenging, especially for non-additive codes that lack a stabilizer description.
We present a general learning-based framework that, given only an encoding circuit, constructs physical implementations of logical operations while enforcing structural properties such as transversality or shallow depth. Our approach is validated by rediscovering known logical operations of standard stabilizer codes. We then extend it to a co-design procedure, dubbed variational early fault-tolerant quantum computing (VarEFTQC), which tailors non-additive encodings to a given noise model and enforces desired logical gate sets, such as transversal IQP-type families or low-depth universal sets.
A software library implements the complete learning pipeline, including loss-function variants, ansatz families, and optimization routines. Together, these results position VarEFTQC as a proof-of-concept framework for discovering hardware-adapted logical gadgets for early fault-tolerant quantum computing.
\end{abstract}

\maketitle

\section{\label{sec:introduction}Introduction}

\Gls{qec} is a prerequisite for scalable quantum computation in the presence of noise and decoherence. Logical qubits are encoded into larger physical Hilbert spaces so that errors can be detected and corrected without directly measuring the logical state of the encoded qubits, and thus without destroying the encoded quantum information~\cite{shor1995scheme,steane1996error}. Beyond algebraically constructed stabilizer codes~\cite{gottesman1997stabilizer}, recent work on noise-tailored and non-additive codes\textemdash{}such as variationally learned encodings~\cite{meyer2025learning,meyer2025variational}\textemdash{}has shown that small, highly specialized codes can outperform traditional stabilizer codes at comparable or smaller qubit counts. However, such learned encodings lack compact algebraic descriptions and do not by themselves provide efficient and fault-tolerant realizations of logical operations.

From the perspective of \gls{ftqc}, however, good encodings alone are not sufficient: one must also implement logical gates that are accurate and structurally compatible with fault-tolerant architectures~\cite{shor1996fault,gottesman2010introduction,aliferis2005quantum,lidar2013quantum}. Transversal implementations are particularly attractive, as their locality natively suppresses error propagation within a code block~\cite{gottesman1997stabilizer,eastin2009restrictions}, but they are constrained both by hardware connectivity\textemdash{}often requiring more elaborate constructions such as lattice surgery~\cite{horsman2012surface}\textemdash{}and by the Eastin–Knill theorem, which rules out a universal set of purely transversal gates for any finite-distance code~\cite{eastin2009restrictions}. In practice, universal logical gate sets therefore rely on additional non-transversal gadgets such as magic-state distillation~\cite{bravyi2005universal} or code switching~\cite{paetznick2013universal}. The recently introduced notion of \gls{eftqc}~\cite{katabarwa2024early} relaxes these requirements, focusing on practical low-depth logical gadgets that substantially mitigate noise under realistic constraints.

\begin{figure}[tb]
    \centering
    \includegraphics{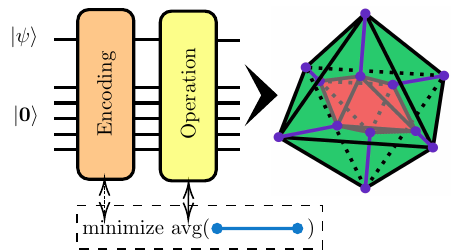}
    \caption{\label{fig:figure1}Schematic overview of the framework for learning logical operations. A pure state $\ket{\psi}$ is encoded into a larger ancilla system using a fixed encoding unitary $U_{\mathrm{enc}}$. Subsequently, a parameterized unitary $U_L(\Psi)$ implements a candidate physical realization of a target logical operation. The parameters are trained with an \emph{operation loss}, measuring the average fidelity over 2-design states between prediction and target (blue lines). Structured operation ans{\"a}tze are used to enforce transversal or low-depth properties. We extend this pipeline to incorporate an adaptable encoding $U_{\mathrm{enc}}(\Theta)$, resulting in a co-design procedure called \gls{vareftqc}.}
\end{figure}

Existing algebraic and algorithmic approaches to constructing logical operations mostly target specific stabilizer codes and typically assume a stabilizer description~\cite{webster2023transversal} or an explicit recovery procedure~\cite{chen2022automated}, making them difficult to apply to arbitrary encodings, in particular non-additive ones. In this work, we address this gap by developing a general variational framework for learning logical operations on \gls{qec} codes specified only through a quantum circuit implementing the encoding unitary $U_{\mathrm{enc}}$, see also \cref{fig:figure1}. The generality claimed here is thus at the level of the required code description\textemdash{}no stabilizer representation, decoder, or explicit recovery circuit is assumed\textemdash{}rather than at the level of present numerical scalability. In its current form, the method relies on explicit classical simulation of the encoded dynamics and is therefore presently restricted to small, classically tractable instances.

Logical gates are implemented by parameterized physical circuits with tunable parameters $\Psi$, which realize unitary $U_L(\Psi)$. A fidelity-based \emph{operation loss} is minimized over 2-designs on the relevant logical subspace so that the resulting physical implementation\textemdash{}the \emph{gadget}\textemdash{}reproduces the desired logical unitary on encoded states. The same machinery applies to intra-block single-logical-qubit gates and inter-block two-logical-qubit gates, and can be instantiated with transversal or shallow non-transversal ansätze. Building on recent work on \gls{varqec}~\cite{meyer2025learning,meyer2025variational}, which learns noise-tailored encodings by minimizing a distinguishability-based loss, we further combine encoding and operation learning into a single co-design framework. This joint optimization, which we term \gls{vareftqc}, simultaneously tailors non-additive codes to a given noise channel and discovers logical gate sets with favorable structural properties for \gls{eftqc}.

\medskip\noindent
\textbf{Main contributions.} In the following, we summarize the main contributions of our work:
\renewcommand{\labelenumi}{\Roman{enumi}.}
\begin{enumerate}
    \item We introduce a variational pipeline that, given only an encoding unitary, learns physical circuit implementations of target logical unitaries. A fidelity-based operation loss over logical 2-designs drives training, and the method applies to single- and multi-qubit logical gates, intra- and inter-block operations, without requiring a stabilizer description, decoder, or explicit recovery circuit, making it applicable to arbitrary codes\textemdash{}including non-additive codes.
    \item We validate the framework by recovering all known transversal single- and two-logical-qubit gates on standard stabilizer codes with one logical qubit per block. We compare diagonal, block-diagonal, and full pairwise operation losses and find that a block-diagonal loss over a 2-design markedly improves optimization stability and success rates. Moderate overparameterization of transversal ansätze via layer repetition further boosts convergence while preserving transversality after compilation.
    \item Combining our procedure with \gls{varqec}~\cite{meyer2025learning}, we co-optimize the encoding and the logical implementations. This \gls{vareftqc} framework discovers small non-additive codes that both suppress information loss for biased noise and admit native logical IQP-type gate sets with transversal structure, and shows that adding a single shallow non-transversal Hadamard gadget suffices to complete a universal logical gate set, consistent with the relaxed, gadget-oriented perspective of \gls{eftqc}.
    \item We implement the learning pipeline, including loss-function variants, ansatz families, and optimization routines, facilitating extensions to other codes, noise models, and hardware architectures; details are given in the data availability statement.
\end{enumerate}
Taken together, these contributions establish a general framework for learning high-fidelity logical operations, extended to the joint co-design of noise-tailored non-additive encodings and logical gate sets. This positions \gls{vareftqc} as a proof-of-concept framework for constructing small, hardware-friendly logical patches and gate gadgets in the spirit of \gls{eftqc}.

\medskip\noindent
The remainder of this paper is organized as follows. \Cref{sec:preliminaries} reviews preliminaries on noise channels, \gls{qec}, \gls{varqec}, logical operations, fault tolerance, and related work. \Cref{sec:method} introduces our framework for learning logical operations from encoders only, including the fidelity-based operation loss, its 2-design evaluation, and transversality-aware ansatz families. In \cref{sec:empirical}, we validate this framework on standard stabilizer codes, studying its ability to (re)discover known transversal gates and the impact of loss variants and ansatz overparameterization. \Cref{sec:vareftqc} presents the \gls{vareftqc} co-design procedure, which jointly optimizes encodings and logical gate sets for biased noise. \Cref{sec:discussion} discusses how these results fit into the broader roadmap towards early and ultimately fully fault-tolerant quantum computing and outlines open challenges, while \cref{sec:summary} summarizes our findings.


\section{\label{sec:preliminaries}Preliminaries and Related Work}

In this section, we summarize the ingredients needed in the remainder of the paper and connect our approach to previous work. We first define the notation for noise channels, then briefly review \gls{qec}, variationally designed encodings, and the notions of logical operations, as well as fault tolerance.

Quantum states of an $n$‑qubit system are described by density operators $\rho$ on a $2^n$‑dimensional Hilbert space, i.e., positive semidefinite operators with $\mathrm{Tr}(\rho)=1$. Pure states are special cases of the form $\rho = \ket{\psi}\!\bra{\psi}$, while mixed states represent classical ensembles of pure states.

General noisy evolutions are modeled by \gls{cptp} maps $\mathcal{N}$, which admit a Kraus representation
\begin{equation}
    \mathcal{N}(\rho) = \sum_k E_k \rho E_k^\dagger,
    \text{ with }
    \sum_k E_k^\dagger E_k = I
\end{equation}
with Kraus operators $E_k$ acting on the system Hilbert space~\cite{kraus1971general}.

In this work, we are mainly concerned with single‑qubit Pauli noise, i.e., channels of the form
\begin{equation}
    \mathcal{N}_{\mathrm{Pauli}}(\rho)
    = (1-p)\rho + p_X X\rho X + p_Y Y\rho Y + p_Z Z\rho Z,
    \label{eq:pauli_noise}
\end{equation}
where $p = p_X + p_Y + p_Z$ is the total error probability and $X,Y,Z$ are the Pauli operators. The special case $p_X=p_Y=p_Z=\frac{p}{3}$ corresponds to the typical symmetric depolarizing noise. For this paper, we also consider \emph{asymmetric depolarizing} noise $\mathcal{N}_{\mathrm{adep}}$, where phase‑flip errors dominate over bit flips, characterized by an asymmetry parameter $c$ that fixes the ratios $\frac{\log p_Z}{\log p_X}=\frac{\log p_Z}{\log p_Y}$~\cite{olle2024simultaneous,meyer2025learning}. Additionally, we also evaluate on simplified \emph{bitflip} noise $\mathcal{N}_{\mathrm{bit}}$ with $p_X=p$ and $p_Y=p_Z=0$.

We assume that errors on different physical qubits are independent, so that the noise on an $n$‑qubit register is modeled as $\mathcal{N}^{\otimes n} = \bigotimes_{j=1}^{n} \mathcal{N}_j$. Extensions to correlated and non‑Pauli channels (e.g., amplitude damping, thermal relaxation) are conceptually possible\textemdash{}as demonstrated in Ref.~\cite{meyer2025learning}\textemdash{}but are not the focus of this manuscript.


\subsection{\label{subsec:preliminaries_qec}Quantum Error Correction}

\Glsentrylong{qec} (QEC) protects logical information by encoding it into a larger physical Hilbert space so that typical noise can be detected and corrected~\cite{shor1995scheme,steane1996error}. A general code $\mathcal{C}$ is characterized by parameters
\begin{equation}
    ((n,K,d)),
\end{equation}
where $n$ is the number of physical qubits, $K$ is the logical Hilbert‑space dimension, and $d$ is the code distance. When $K=2^k$ for some integer $k$, the code encodes $k$ logical qubits. 
A distance-$d$ code can detect any error acting nontrivially on up to $d-1$ qubits and can correct up to $\lfloor (d-1)/2 \rfloor$ arbitrary qubit errors. For codes without a proven distance, we simply write $((n,K))$.
For stabilizer (\emph{additive}) codes~\cite{gottesman1997stabilizer} we use the standard notation
\begin{equation}
    [[n,k,d]],
\end{equation}
where $k=\log_2 K$ is the number of logical qubits. In this paper we consider both standard stabilizer codes such as the $[[3,1,1]]$ \emph{bitflip}, $[[4,1,2]]$ \emph{approximate}, $[[5,1,3]]$ \emph{perfect}, $[[7,1,3]]$ \emph{Steane}, $[[9,1,3]]$ \emph{Shor}, $[[10,1,4]]$ \emph{short dodeca}, $[[11,1,5]]$ \emph{dodeca}, and $[[15,1,3]]$ \emph{Reed-Muller} codes~\cite{shor1995scheme,leung1997approximate,laflamme1996perfect,steane1996error,olle2024simultaneous,steane2002quantum}, as well as non‑additive variational codes~\cite{meyer2025learning,meyer2025variational} (see \cref{subsec:preliminaries_VarQEC}).

Given a code $\mathcal{C} \subset \mathcal{H}^{\otimes n}$ with $k$ logical qubits, an encoding unitary $U_{\mathrm{enc}}$ maps a $k$‑qubit input state $\rho$ and $n-k$ ancilla qubits in $\ket{0}$ to an encoded state
\begin{equation}
    \rho_L = U_{\mathrm{enc}}\bigl(\rho \otimes \ket{0}\!\bra{0}^{\otimes n-k}\bigr)U_{\mathrm{enc}}^\dagger.
    \label{eq:encoding}
\end{equation}
The restriction of $U_{\mathrm{enc}}$ to states of the form $\rho \otimes \ket{0}^{\otimes n-k}$ defines an isometric embedding of the logical space into the physical $n$-qubit space, and its image is the codespace $\mathcal{C}$. In a full \gls{qec} cycle, the encoded state passes through a noise channel $\mathcal{N}$, then potentially multiple cycles of a recovery operation, and finally a decoding step $U_{\mathrm{enc}}^\dagger$ returns the state to the original $k$‑qubit subsystem by additionally tracing out the ancillary degrees of freedom.


\subsection{\label{subsec:preliminaries_VarQEC}Variational Code Design}

Standard codes such as the surface code~\cite{fowler2012surface} or the $[[5,1,3]]$ \emph{perfect} code are designed to handle generic noise patterns, often at the expense of significant overhead. For near- and mid‑term devices, it can be advantageous to tailor the encoding to the dominant noise channel. \Glsentrylong{varqec} addresses this by learning the encoding from data about the noise~\cite{meyer2025variational,meyer2025learning}. We note that historically the acronym ``VarQEC'' was first introduced by Cao et al.~\cite{cao2022quantum} for a variational, Knill–Laflamme‑based, channel‑adaptive code‑discovery framework. In this work, unless explicitly stated otherwise, VarQEC refers instead to the trace‑distance‑based, distinguishability‑loss approach described below~\cite{meyer2025learning}, and not to the Knill–Laflamme‑based formulation~\cite{cao2022quantum}.

The central idea is to measure how much \emph{distinguishability} between quantum states is lost under encoding and noise. Given an encoding $U_{\mathrm{enc}}$, optionally parameterized by $\Theta$, and a noise channel $\mathcal{N}$, the \emph{lost trace distance} is
\begin{equation}
    \Delta_T(\rho,\sigma;\mathcal{N};\Theta)
    = T(\rho,\sigma) - T\bigl(\mathcal{N}(\rho_L),\mathcal{N}(\sigma_L)\bigr),
\end{equation}
where $\rho_L$ and $\sigma_L$ are encoded according to Eq.~\eqref{eq:encoding}, and $T$ is the standard trace distance between pure states~\cite{watrous2018theory}. The corresponding \emph{worst‑case distinguishability loss}
\begin{equation}
    \overline{\mathcal{D}}(\mathcal{N};\Theta)
    = \max_{\rho,\sigma} \Delta_T(\rho,\sigma;\mathcal{N};\Theta)
\end{equation}
is small if the encoding preserves distinguishability for all logical states, which in turn guarantees the existence of a high‑fidelity recovery operation~\cite{meyer2025variational,johnson2017qvector}.
Directly optimizing the worst‑case expression is unstable and computationally expensive, so \gls{varqec} uses an \emph{average‑case proxy} over a two‑design $\mathcal{S}$:
\begin{equation}
    \mathcal{D}_S(\mathcal{N};\Theta)
    = \frac{1}{|\mathcal{S}|^2}
      \sum_{\rho,\sigma\in \mathcal{S}}
      \Delta_T(\rho,\sigma;\mathcal{N};\Theta).
\end{equation}
In \gls{varqec}, $U_{\mathrm{enc}}(\Theta)$ is instantiated as a parameterized circuit, typically a \gls{rea} with single‑qubit rotations and randomly placed two‑qubit gates constrained by hardware connectivity~\cite{meyer2025learning}. The parameters are then optimized classically via
\begin{equation}
    \min_{\Theta}\, \mathcal{D}_S(\mathcal{N};\Theta).
\end{equation}
For several Pauli but also non-unital noise structures, this yields small non‑additive codes $((n,2))$ that match or surpass the performance of traditional stabilizer‑type codes at comparable or smaller $n$~\cite{meyer2025variational,meyer2025learning}. In Ref.~\cite{meyer2025learning}, such \gls{varqec} encodings were further combined via noise‑aware concatenation to suppress logical error rates over multiple levels. On structured noise, this yields significant overhead reductions compared to the concatenation of just standard stabilizer codes.


\subsection{\label{subsec:preliminaries_operations}Logical Operations and Fault-Tolerance}

Once a code $\mathcal{C}$ and its encoding unitary $U_{\mathrm{enc}}$ are fixed, logical density operators $\rho$ on $k$ logical qubits are mapped to encoded states $\rho_L$ via \cref{eq:encoding}. A \emph{logical operation} is a unitary $U$ acting on the logical $k$‑qubit space that is implemented physically by an $n$‑qubit unitary $U_L$ on the code block. In the ideal, exact case, one requires
\begin{equation}
    \label{eq:logical_gate_exact}
    U_L \,\rho_L\, U_L^\dagger
    = (U \rho U^\dagger)_L
\end{equation}
for all logical states $\rho$, with the convenient notation $(U \rho U^\dagger)_L = U_{\mathrm{enc}}\,(U\rho U^\dagger)\,U_{\mathrm{enc}}^\dagger$. Formulated slightly differently, this means that $U_L U_{\mathrm{enc}}$ and $U_{\mathrm{enc}} U$ agree on the data‑plus‑ancilla subspace up to a global phase~\cite{nielsen2010quantum,gottesman1997stabilizer,lidar2013quantum}.
In practice, implementations are only approximate, and we write
\begin{equation}
    \label{eq:logical_gate_approximate}
    U_L \,\rho_L\, U_L^\dagger \approx (U \rho U^\dagger)_L
\end{equation}
to indicate that equality holds up to a small error on encoded states. More formally, this approximate equality can be understood in terms of the distance between the corresponding logical channels, for example, by requiring that the completely positive trace‑preserving map induced by $U_L$ on the codespace be close in diamond norm to the ideal logical channel induced by $U$~\cite{watrous2018theory}. In this work, we do not estimate such channel distances directly, but instead quantify the error via state fidelities and the operation‑loss metrics introduced in~\cref{sec:method}, in line with standard notions of approximate logical gates~\cite{aliferis2005quantum,gottesman2010introduction}. Throughout, we denote target single- and two-logical-qubit gates by elements of $\mathrm{U}(2)$ and $\mathrm{U}(4)$, respectively, understood up to physically irrelevant global phase.

A central requirement in \gls{ftqc} is that logical operations must be implemented so that a small number of physical faults during the gadget do not create more errors on any code block than the code can correct. Roughly, for a code that corrects up to $t$ errors per block, a logical gate implementation is called fault‑tolerant if $r$ faults inside the gadget lead to at most $\mathcal{O}(r)$ physical errors on each block, so that the resulting state is still within the correctable set~\cite{shor1996fault,gottesman2010introduction,aliferis2005quantum,lidar2013quantum}.

Within this landscape, a particularly simple and widely used structural property is \emph{transversality}~\cite{shor1996fault,gottesman1997stabilizer,eastin2009restrictions}. Intuitively, a transversal logical gate acts `qubit‑wise' across the code: for a logical operation within one code block (\emph{intra-block}) it is implemented by a layer of single‑qubit gates on the physical qubits of that block, and for a logical gate between two code blocks (\emph{inter-block}) it is implemented by two-qubit gates, each acting on exactly one qubit from each block. Such locality constraints ensure that a single physical fault during the gate can create at most one error per block and therefore naturally support fault tolerance. At the same time, the Eastin–Knill theorem shows that no finite‑distance code admits a universal set of purely transversal logical gates, necessitating the use of additional, non‑transversal but still fault‑tolerant gadgets in universal \gls{ftqc} architectures~\cite{eastin2009restrictions}.

From the perspective of \gls{eftqc}~\cite{katabarwa2024early}, however, it can already be valuable to employ logical implementations that substantially reduce depth or overhead even if they do not yet form a universal, fully fault‑tolerant gate set. In this regime, transversality is best viewed as a desirable structural property and a proxy for hardware‑friendliness, rather than a strict requirement that every logical operation must satisfy.

For this work, we can treat standard stabilizer codes and learned encodings (see \cref{subsec:preliminaries_VarQEC}) the same way: any code for which we know $U_{\mathrm{enc}}$ is a valid input to our logical‑operation learning procedure. Furthermore, in \cref{sec:vareftqc}, we extend upon this by reusing $\mathcal{D}_S$ as part of a joint loss that co‑optimizes the encoding and selected logical gates.


\subsection{\label{subsec:preliminaries_related}Related Work on Constructing Operations}

Logical operations for quantum error‑correcting codes have been studied extensively, particularly for stabilizer codes. Many small‑distance codes admit hand‑designed constructions of logical Clifford gates and sometimes non‑Clifford gates. Examples include transversal or blockwise implementations of Clifford gates for the $[[7,1,3]]$ Steane code and the $[[9,1,3]]$ Shor code~\cite{steane1996error,shor1995scheme}, and transversal $T$ gates for the $[[15,1,3]]$ Reed–Muller code used in magic‑state distillation~\cite{gottesman1997stabilizer}. In surface‑code architectures, logical gates are typically implemented via braiding, lattice surgery, or code deformation rather than finite‑depth transversal circuits~\cite{fowler2012surface}.

Beyond such analytic constructions, algorithmic methods have been proposed to automate the search for logical gates. On the algorithmic side, Webster et al.~\cite{webster2023transversal} use the XP stabilizer formalism to efficiently characterize transversal diagonal logical operators on CSS and more general stabilizer codes, providing scalable tools to enumerate depth‑one diagonal logical gates at arbitrary levels of the Clifford hierarchy, with refinements of the procedure proposed in Ref.~\cite{camps2026transversal}. In cases of low-distance stabilizer codes like $[[n,n-2,2]]$ \emph{Iceberg} codes, there are also algorithmic constructions that efficiently compile entire Clifford sub-circuits, instead of resorting to typical gate-by-gate realization~\cite{chen2025tailoring,popov2026optimized,chen2025fault}.

On the learning side, Chen et al.~\cite{chen2022automated} introduced a variational procedure that is conceptually close to ours: both the logical gate and its physical implementation are represented by parameterized circuits, and a loss over a tomographically complete set of logical input states is minimized so that encoding, physical ansatz, error correction, and decoding jointly reproduce the desired logical operation. Their method successfully recovers and discovers logical gates for a range of small stabilizer codes, but relies on detailed prior knowledge of a recovery procedure (syndrome measurements plus minimum‑weight correction) and on process tomography over a large set of logical states, which severely limits scalability and flexibility.

Our approach shares with Chen et al.~\cite{chen2022automated} the use of variational circuits and a fidelity‑based objective, but differs from existing work in several fundamental ways. Most importantly, we only assume access to an encoding unitary $U_{\mathrm{enc}}$ and do not require an explicit recovery or decoding circuit, which makes the method applicable to arbitrary codes, including non‑additive \gls{varqec} encodings without a convenient stabilizer description. Additionally, our framework uniformly treats single‑ and two‑logical‑qubit gates, intra‑ and inter‑block operations, and both transversal and non‑transversal realizations. Finally, we combine learning operations with the tailoring of the encoding itself, allowing us to jointly optimize encodings and logical gate sets for a given noise model, a capability not present in previous works.


\section{\label{sec:method}Learning Logical Operations}

Given any \gls{qec} code $\mathcal{C}$ for which we have access to its encoding unitary $U_{\mathrm{enc}}$, this section introduces a procedure to learn physical implementations of logical unitaries acting on the codespace. We deliberately do not assume access to an explicit decoding or recovery circuit, nor to a stabilizer description of the code, which makes the framework applicable to arbitrary codes. The central idea is to quantify how well a parameterized ansatz $U_L(\Psi)$ approximates a desired logical unitary $U$ in the sense of \cref{eq:logical_gate_approximate}, and to optimize its parameters $\Psi$ variationally.

The framework is, in principle, applicable to codes with an arbitrary number of logical qubits $k$ and to logical unitaries acting on any subset of those logical qubits, possibly spread across several blocks. The only code-specific ingredient is $U_{\mathrm{enc}}$. However, for this first proof-of-concept realization, we restrict our explicit formulations to setups with $k=1$. In particular, we discuss how to learn single-qubit intra-block operations in \cref{subsec:method_single_qubit}, and extend this to two-qubit inter-block operations in \cref{subsec:method_multi_qubit}. Given a suitable set of intra-block logical operations such as $\{H, T\}$ and inter-block logical operations such as $\{\mathrm{CNOT}\}$, this already suffices for universal computation. Navigating the loss landscape with undesirable local optima requires modifications to the naive loss function, as discussed in \cref{subsec:method_loss}, while ansatz choices and their implications for fault tolerance are addressed in \cref{subsec:method_transversality}.

\subsection{\label{subsec:method_single_qubit}Learning Intra-Block Operations}

\begin{figure*}[tb]
    \centering
    \includegraphics{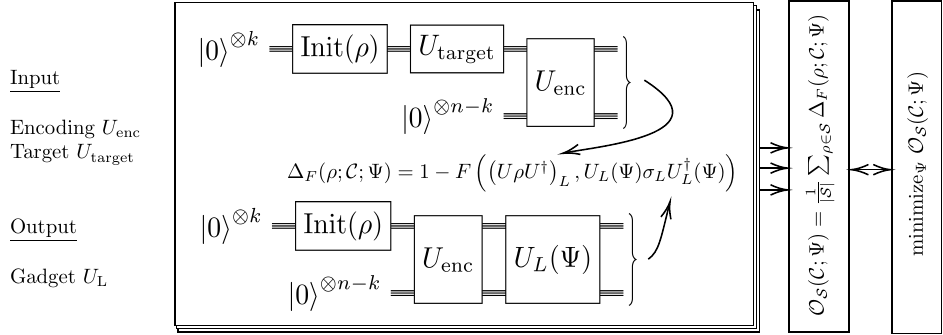}
    \caption{\label{fig:pipeline}Variational learning pipeline for a logical operation $U$ on a fixed code $C$ using only the encoding unitary $U_{\mathrm{enc}}$. Logical input states are drawn from a 2-design $\mathcal{S}$ on the relevant logical subsystem, encoded, and passed through the parameterized prediction $U_L(\Psi)$. The resulting encoded states are compared to the target encoded states $(U \rho U^{\dagger})_L$ using the fidelity-based operation error $\Delta_F$, and the parameters $\Psi$ are optimized to minimize the average-case operation loss $\mathcal{O}_{\mathcal{S}}(C; \Psi)$.}
\end{figure*}

Let $\rho$ be a pure logical state on $k$ qubits, and let $\rho_L$ denote its encoded version according to \cref{eq:encoding}. The ideal encoded output of $U$ is $(U \rho U^{\dagger})_L$, obtained by applying $U$ before encoding. The physical implementation applies the ansatz $U_L(\Psi)$ after encoding, resulting in $U_L(\Psi) \rho_L U_{L}^{\dagger}(\Psi)$. We measure the closeness of pure states by the state fidelity $F(\rho,\sigma)$, which is used to define a measure for the quality of candidate implementations $U_L(\Psi)$  for a fixed logical input state $\rho_L$ by the \emph{operation error}
\begin{align}
    \label{eq:operation_error}
    \Delta_{F}(\rho;\mathcal{C};\Psi) = 1 - F((U \rho U^{\dagger})_L, U_L(\Psi) \rho_L U_{L}^{\dagger}(\Psi)).
\end{align}
By construction, $\Delta_F(\rho; C; \Psi) = 0$ if and only if \cref{eq:logical_gate_exact} holds for the input $\rho$. Based on this, we define the \emph{worst-case operation loss}
\begin{align}
    \label{eq:operation_loss_worst_case}
    \overline{\mathcal{O}}(\mathcal{C};\Psi) = \max_{\rho} \Delta_{F}(\rho;\mathcal{C};\Psi),
\end{align}
where the maximization is over all pure logical input states $\rho$ on the subsystem on which $U$ acts. It is known from classical machine learning that training with loss functions that incorporate extrema operations can be quite unstable~\cite{hastie2009elements,bassily2020stability}, which we also observed here. Therefore, we introduce the respective \emph{average-case operation loss} as
\begin{align}
    \label{eq:operation_loss_average_case}
    \mathcal{O}(\mathcal{C};\Psi) = \int_{\rho} \Delta_{F}(\rho;\mathcal{C};\Psi) \,d\nu(\rho),
\end{align}
with $d\nu$ denoting the Haar measure on the logical subsystem~\cite{sommers2004statistical}.
Trivially, $\mathcal{O}(C; \Psi) \le \overline{\mathcal{O}}(C; \Psi)$, but also $\mathcal{O}(C; \Psi) = 0 \Leftrightarrow \overline{\mathcal{O}}(C; \Psi) = 0$, i.e., the average-case version is a reasonable proxy for the worst-case formulation.

Throughout most of this work, we assume an arbitrary target logical unitary $U$ and omit it from the notation. When it is not clear from context, we explicitly indicate the respective operation $U$ by the notation 
\begin{align}
    \overline{\mathcal{O}}^{U}(\mathcal{C};\Psi^U) \text{ ~or~ } \mathcal{O}^{U}(\mathcal{C};\Psi^U),
\end{align}
in particular in cases where multiple operations are considered in one expression.

To reduce the computational cost of evaluating \cref{eq:operation_loss_average_case}, we replace the Haar average over pure logical input states by an average over a finite exact complex projective 2-design on the logical Hilbert space on which $U$ acts~\cite{klappenecker2005mutually,renes2004symmetric}. For the analogous notion defined on unitary operators, see~\cite{ambainis2007quantum,dankert2009exact}; here, however, we work with state 2-designs. Let $\mathcal{S}$ denote such a design, with each element identified with the rank-one projector $\rho=\ket{\psi}\!\bra{\psi}$. We then define the \emph{2-design operation loss} as
\begin{align}
    \label{eq:operation_loss_average_case_two_design}
    \mathcal{O}_{\mathcal{S}}(\mathcal{C};\Psi) = \frac{1}{\abs{\mathcal{S}}} \sum_{\rho \in \mathcal{S}} \Delta_{F}(\rho;\mathcal{C};\Psi).
\end{align}
For $\rho=\ket{\psi}\!\bra{\psi}$, the quantity $\Delta_F(\rho;\mathcal{C};\Psi)$ is a polynomial of bidegree $(2,2)$ in the amplitudes of $\ket{\psi}$ and their complex conjugates. By the defining property of a complex projective 2-design, averaging over $\mathcal{S}$ therefore reproduces the Haar average exactly, so that
\begin{align}
    \mathcal{O}(\mathcal{C}; \Psi)
    =
    \mathcal{O}_{\mathcal{S}}(\mathcal{C}; \Psi).
\end{align}
Hence, replacing the Haar integral by the finite average over $\mathcal{S}$ introduces no approximation at the level of the loss.

\bigskip\noindent
Given a fixed code $C$, a target logical unitary $U$, and an ansatz family $U_L(\Psi)$, we search for a realization of the logical operation by minimizing the 2-design loss with respect to $\Psi$:
\begin{align}
    \label{eq:training_minimization}
    \min_{\Psi}\;
        \mathcal{O}_{\mathcal{S}}(C; \Psi).
\end{align}
This defines the basic variational pipeline shown in \cref{fig:pipeline}: logical input states are sampled from $\mathcal{S}$, encoded by $U_{\mathrm{enc}}$, evolved under the \emph{prediction} $U_L(\Psi)$, and compared with the \emph{target} encoded outputs using the fidelity in \cref{eq:operation_error}.
We declare a logical implementation $U_L(\Psi)$ to have \emph{quality} $\varepsilon$ if its worst-case error over the 2-design states is at most $\varepsilon$, i.e., $\overline{\mathcal{O}}_{\mathcal{S}}(C; \Psi) \le \varepsilon$. Throughout this work, we consider a logical operation to be successfully realized if $\varepsilon \le 10^{-5}$.

\bigskip\noindent
For a single logical qubit, i.e., $k=1$, and intra-block logical gates $U \in \mathrm{U}(2)$, understood up to physically irrelevant global phase, we choose the state 2-design
\begin{align}
    \mathcal{S}^{(1)} = \mathcal{B}_Z \cup \mathcal{B}_X \cup \mathcal{B}_Y,
\end{align}
where each state is again identified with the corresponding rank-one projector, and
\begin{align}
    \label{eq:mub_single_qubit}
    \mathcal{B}_Z &= \left\{ \ket{0}, \ket{1} \right\}, \nonumber\\
    \mathcal{B}_X &= \left\{ \ket{+}, \ket{-} \right\}, \\
    \mathcal{B}_Y &= \left\{ \ket{+i}, \ket{-i} \right\}. \nonumber
\end{align}
These are the eigenbases of the Pauli $Z$, $X$, and $Y$ operators. In dimension $d=2$, they form a complete set of $d+1=3$ \gls{mub}, and the union of a complete set of \glspl{mub} is an exact complex projective 2-design~\cite{klappenecker2005mutually}. An alternative exact choice in $d=2$ is a \gls{sicpovm}, i.e., four pure states with pairwise overlaps $|\langle \phi_j | \phi_k \rangle|^2 = 1/3$ for $j \neq k$~\cite{renes2004symmetric}. In this work, however, we adopt the \gls{mub}-based construction because it is simpler to implement and experimentally more accessible.


\subsection{\label{subsec:method_multi_qubit}Learning Inter-Block Operations}

\begin{figure}[tb]
    \centering
    \subfigure[\label{subfig:target_two_qubit}Target: logical inter-block operation.]{
        \includegraphics{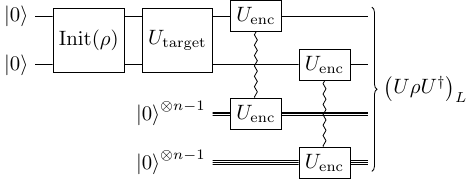}
    }\\
    \subfigure[\label{subfig:prediction_two_qubit}Prediction: Physical inter-block operation.]{
        \includegraphics{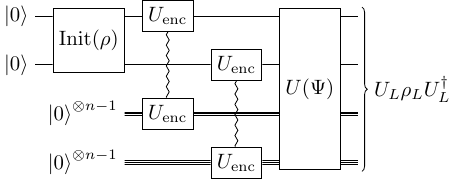}
    }
    \caption{\label{fig:models_two_qubit}Circuit-level view of the learning setup for a two-qubit logical gate $U\in\mathrm{U}(4)$ acting on two encoded blocks of the same code $C$: (a) the target operation is applied to the logical state before encoding; (b) the variational prediction first encodes both logical qubits and then applies a parameterized ansatz on the $2n$ physical qubits. The resulting predicted state is compared to the target one using the operation error defined in \cref{eq:operation_error}.}
\end{figure}

Extending the proposed pipeline to support multi-qubit operations between code blocks requires only small modifications: let $\mathcal{C}$ be a code with $k$ logical qubits per block and encoding $U_{\mathrm{enc}}$. Two blocks then encode a two-qubit logical state $\rho$ on $2k$ qubits as
\begin{align}
    \label{eq:encoding_two_qubit}
    \rho_L = U_{\mathrm{enc}}^{\otimes 2} \left( \rho \otimes \ket{0} \bra{0}^{\otimes 2(n-k)} \right) U_{\mathrm{enc}}^{\otimes 2\dagger},
\end{align}
where $\rho$ may be entangled across the two logical subsystems. Here, \(U_{\mathrm{enc}}^{\otimes 2}\) is understood to act blockwise on the two encoded blocks, i.e., each copy of \(U_{\mathrm{enc}}\) acts on one \(k\)-qubit data register together with its corresponding \(n-k\) ancillas. The obvious regrouping of registers needed to pass from the written ordering to \((\text{data},\text{ancilla})\otimes(\text{data},\text{ancilla})\) is suppressed for readability. A target logical gate $U$ acts on $\rho$ before encoding, while a parameterized physical prediction $U_L(\Psi)$ acts on the $2n$ physical qubits of the two blocks after encoding. This procedure is sketched for $k=1$, i.e., a target two-qubit gate $U \in \mathrm{U}(4)$ between the two blocks, in \cref{fig:models_two_qubit}.

The definitions of operation error and loss carry over trivially: one simply replaces $\rho$ by a $2k$-qubit logical state and uses the encoding from \cref{eq:encoding_two_qubit}. In particular, \cref{eq:operation_error,eq:operation_loss_worst_case,eq:operation_loss_average_case,eq:operation_loss_average_case_two_design} apply without modification.

\bigskip\noindent
For two logical qubits and target gates $U \in \mathrm{U}(4)$, again understood up to physically irrelevant global phase, we choose the state 2-design $\mathcal{S}^{(2)}$ as the union of a complete set of five mutually unbiased bases in dimension $d=4$,
\begin{align}
    \label{eq:two_design_two_qubit}
    \begin{split}
        \mathcal{S}^{(2)}
        ={}&
        (\mathcal{B}_{ZZ})
        \cup
        (\mathcal{B}_{XX})
        \cup
        (\mathcal{B}_{YY}) \\
        &\cup\,
        CZ(\mathcal{B}_{XY})
        \cup
        CZ(\mathcal{B}_{YX}),
    \end{split}
\end{align}
where each state is again identified with the corresponding rank-one projector. Here,
\begin{align}
    \mathcal{B}_{\alpha \beta}
    =
    \left\{
        \ket{\phi}\otimes\ket{\chi}
        \,:\,
        \ket{\phi}\in\mathcal{B}_\alpha,\;
        \ket{\chi}\in\mathcal{B}_\beta
    \right\},
\end{align}
with $\mathcal{B}_Z$, $\mathcal{B}_X$, and $\mathcal{B}_Y$ defined as in \cref{eq:mub_single_qubit}, while $CZ(\mathcal{B}_{\alpha \beta})$ denotes the basis obtained by applying the controlled-$Z$ gate to each state in $\mathcal{B}_{\alpha \beta}$. We adopt the explicit \gls{mub}-based form of $\mathcal{S}^{(2)}$ in \cref{eq:two_design_two_qubit} because it permits streamlined implementation and is experimentally more accessible, although a smaller exact state 2-design with only $d^2=16$ states could alternatively be constructed from a \gls{sicpovm} in dimension $d=4$~\cite{renes2004symmetric}. As the union of a complete set of \glspl{mub} in dimension $d=4$, $\mathcal{S}^{(2)}$ is an exact complex projective 2-design~\cite{klappenecker2005mutually}; it therefore contains $d(d+1)=20$ elements and reproduces the Haar-average loss exactly.

\bigskip\noindent
More generally, the same idea extends directly to target operations acting on $m$ logical qubits, irrespective of whether these qubits lie within a single block or are distributed across several blocks. One may choose $\mathcal{S}^{(m)}$ as any exact complex projective 2-design on the corresponding logical Hilbert space of dimension $d=2^m$. Since $d$ is a prime power, a convenient exact choice is again the union of a complete set of $d+1$ mutually unbiased bases, which yields $d(d+1)=2^m(2^m+1)=\mathcal{O}(4^m)$ design elements~\cite{klappenecker2005mutually}. This still reproduces the Haar-average loss exactly, but the number of design elements grows exponentially with $m$, so in practice we restrict explicit constructions and numerical experiments to the one- and two-qubit cases. Nevertheless, as discussed in \cref{subsec:discussion_challenges}, patch-based logical architectures provide an alternative route toward scaling to arbitrarily large computations.


\subsection{\label{subsec:method_loss}Variants of Operation Loss}

\begin{figure}[htb!]
    \centering
    \subfigure[\label{subfig:operation_loss_baseline}Target 2-design states on a single logical qubit.]{
        \includegraphics{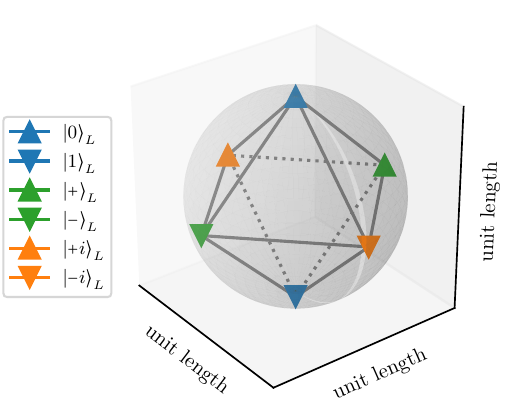}
    }\\
        \subfigure[\label{subfig:operation_loss_diag}Objective for $\mathcal{P}_{\mathrm{diag}}$: minimize length of colored lines.]{
        \includegraphics{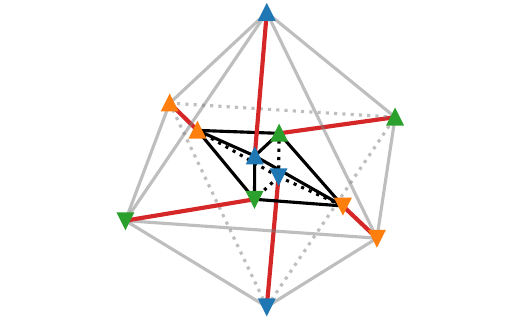}
    }\\
    \subfigure[\label{subfig:operation_loss_block}Objective for $\mathcal{P}_{\mathrm{block}}$: maximize length of colored lines.]{
        \includegraphics{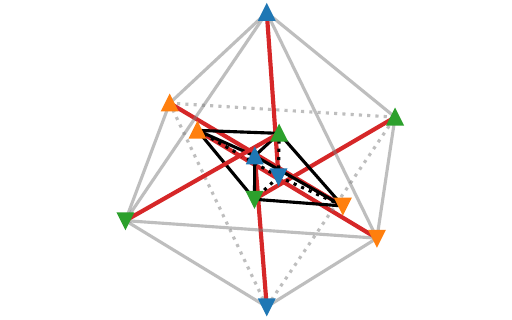}
    }
    \caption{\label{fig:operation_loss_methods}Illustration of different loss variants on the target 2-design octahedron (shown in gray), embedded in the Bloch sphere (a). The diagonal variant in (b) `pulls' the physical prediction (black octahedron) towards the logical target for each 2-design state. The block-diagonal extension in (c) additionally `repels' physical predictions from logical targets that are to be orthogonal. The third variant of evaluating on the full set of state pairs is not explicitly shown.}
\end{figure}

Conceptually, the pipeline introduced in the previous sections already enables learning physical realizations of arbitrary logical operations, both intra- and inter-block. However, we observed that training with the naive loss function in \cref{eq:operation_loss_average_case_two_design} is often unstable and often fails to converge to a loss value in the range of or below $10^{-5}$. We attribute this mostly to the existence of poor local minima in the loss landscape, as we discuss in additional detail in \cref{app:loss}. Therefore, we augment the loss by comparing pairs of logical states, rather than only individual states: the loss now constrains not only each logical state individually but also their relative geometry (overlaps and orthogonality), which empirically removes many of these local minima (see \cref{app:geometric} for a detailed geometric discussion).

For pure states $\rho$ and $\sigma$ on the appropriate logical $k$-qubit subsystems we redefine
\begin{align}
    \begin{split}
        \label{eq:operation_error_extended}
        \Delta_{F}(\rho,\sigma;\mathcal{C};\Psi) =& \bigl| F(U \rho U^{\dagger}, U \sigma U^{\dagger})
        \\&- F((U \rho U^{\dagger})_L, U_L(\Psi)\sigma_L U_{L}^{\dagger}(\Psi)) \bigr|,
    \end{split}
\end{align}
where both encoded states are obtained from \cref{eq:encoding} or \cref{eq:encoding_two_qubit}, depending on if we target an intra- or inter-block operation. We use the absolute difference, rather than a squared discrepancy, since we empirically found it to be more stable during optimization. For identical states $\rho = \sigma$, \cref{eq:operation_error_extended} trivially reduces to \cref{eq:operation_error}.

Given a finite set $\mathcal{P}$ of such state pairs, we define the pairwise worst-case operation loss
\begin{align}
    \overline{\mathcal{O}}_{\mathcal{P}}(\mathcal{C};\Psi) = \max_{(\rho, \sigma) \in \mathcal{P}} \Delta_{F}(\rho,\sigma;\mathcal{C};\Psi),
\end{align}
and the corresponding average-case version
\begin{align}
    \mathcal{O}_{\mathcal{P}}(\mathcal{C};\Psi) = \frac{1}{\abs{\mathcal{P}}} \sum_{(\rho, \sigma) \in \mathcal{P}} \Delta_{F}(\rho,\sigma;\mathcal{C};\Psi).
\end{align}
The choice of $\mathcal{P}$ controls how much \emph{geometric information} (for details on the geometric interpretation of the loss functions see \cref{app:geometric}) about the logical state space is enforced during training. In this work, we consider three natural choices, all constructed from the
same underlying logical 2-design $\mathcal{S}$:

\medskip\noindent
First, we only compare each state's prediction to its target, which can also be interpreted as comparing elements \emph{along the diagonal} of state combinations. This reads as
\begin{align}
    \mathcal{P}_{\mathrm{diag}} = \left\{ (\rho, \rho) \mid \rho \in \mathcal{S} \right\},
\end{align}
which coincides with the direct 2-design formulation as argued above, i.e., $\mathcal{O}_{\mathcal{S}}(\mathcal{C};\Psi) = \mathcal{O}_{\mathrm{diag}}(\mathcal{C};\Psi)$.

Second, we additionally compare mutually orthogonal states in $\mathcal{S}$, i.e., also include \emph{block-diagonal} elements into our evaluation. This results in
\begin{align}
    \mathcal{P}_{\mathrm{block}} = \mathcal{P}_{\mathrm{diag}} \cup \left\{ (\rho,\sigma) \mid \rho \perp \sigma; \rho, \sigma \in \mathcal{S} \right\},
\end{align}
thereby encouraging the physical implementation to preserve not only individual states but also orthogonality relations.

Third, we compare on the \emph{full} set of ordered pairs by
\begin{align}
    \mathcal{P}_{\mathrm{full}} = \left\{ (\rho,\sigma) \mid \rho, \sigma \in \mathcal{S} \right\}.
\end{align}
This loss enforces that the entire Gram matrix of pairwise fidelities between the 2-design states is preserved as well as possible by the variational physical implementation. Information‑theoretically, this approximately preserves all pairwise order-$\nicefrac{1}{2}$ R{\'e}nyi divergences within the logical two‑design ensemble~\cite{watrous2018theory}, so that the gadget acts nearly isometrically on the corresponding logical subspace and does not significantly reduce the amount of classical information that can be encoded and recovered using these logical states.

\medskip\noindent
We also schematically visualize the effect of $\mathcal{P}_{\mathrm{diag}}$ and $\mathcal{P}_{\mathrm{block}}$ on the 2-design octahedron for single-qubit logical operations in \cref{fig:operation_loss_methods}. Since $\mathcal{P}_{\mathrm{diag}} \subseteq \mathcal{P}_{\mathrm{block}}\subseteq \mathcal{P}_{\mathrm{full}}$, the corresponding worst-case losses obey
\begin{align}
    \label{eq:worst_case_bounds}
    \overline{\mathcal{O}}_{\mathrm{diag}}(\mathcal{C};\Psi) \leq \overline{\mathcal{O}}_{\mathrm{block}}(\mathcal{C};\Psi) \leq \overline{\mathcal{O}}_{\mathrm{full}}(\mathcal{C};\Psi).
\end{align}
Geometrically, going from $\mathcal{P}_{\mathrm{diag}}$ to $\mathcal{P}_{\mathrm{block}}$ and $\mathcal{P}_{\mathrm{full}}$ progressively constrains the relative arrangement of the encoded logical states on the Bloch sphere, or its higher-dimensional generalization.

In practice, we always optimize an average-case loss of the form $\mathcal{O}_{\mathcal{P}}(C; \Psi)$ with a suitable choice of $\mathcal{P}$, i.e.,
\begin{align}
    \label{eq:training_minimization_pariwise}
    \min_{\Psi} \mathcal{O}_{\mathcal{P}}(\mathcal{C};\Psi).
\end{align}
In \cref{sec:empirical}, we demonstrate that in most cases $\mathcal{P}_{\mathrm{block}}$ leads to the most stable training behavior. After convergence, the average-case formulation enforces tight bounds on the worst-case variant $\overline{\mathcal{O}}_{\mathcal{P}}(\mathcal{C}; \Psi)$. By \cref{eq:worst_case_bounds}, all loss variants upper-bound $\overline{\mathcal{O}}_{\mathrm{diag}}(\mathcal{C};\Psi) = \overline{\mathcal{O}}_{\mathcal{S}}(\mathcal{C};\Psi)$, which is used to quantify the $\varepsilon$-quality throughout this work, i.e., the threshold such that $\overline{\mathcal{O}}_{\mathcal{S}}(\mathcal{C};\Psi) \leq \varepsilon$.


\subsection{\label{subsec:method_transversality}Ansatz Family and Transversality}

\begin{figure}[tb]
    \centering
    \subfigure[\label{subfig:ansatz_one_qubit}Ansatz for intra-block one-qubit operation.]{
        \includegraphics[width=0.33\textwidth]{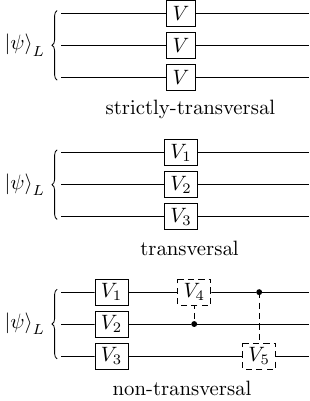}
    }\\
    \subfigure[\label{subfig:ansatz_two_qubit}Ansatz for inter-block two-qubit operation.]{
        \includegraphics[width=0.33\textwidth]{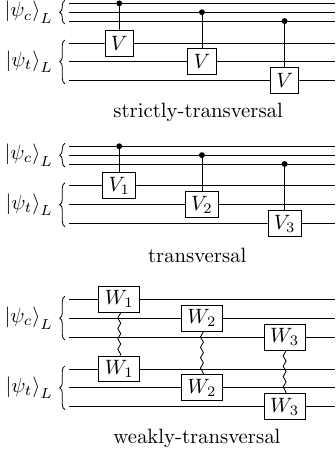}
    }
    \caption{\label{fig:ansatz}Ansatz families for learning logical gates. (a) For single-logical-qubit gates, strict transversality uses the same one-qubit gate $V$ on all physical qubits, transversality allows independent gates $V_i$, and non-transversal ans\"atze add a few randomly placed controlled two-qubit gates. (b) For controlled logical gates between two code blocks, strict transversality uses the same controlled block $CV$ on corresponding qubit pairs, transversality allows independent blocks $CV_i$ with fixed control-target order, and weak transversality uses arbitrary two-qubit blocks $W_i$.}
\end{figure}

Up to this point, the framework treats $U_L(\Psi)$ as a generic parameterized circuit, which, in principle, can be any hardware-compatible ansatz. From a practical perspective, however, the physical realizations should respect structural constraints like transversality, to guarantee inherent fault tolerance for \gls{ftqc}, or at least shallow circuit depth for \gls{eftqc}. In the following, we state the different ansatz families and transversality types considered in this work. Throughout, we assume codes $\mathcal{C}$ with $n$ physical qubits and a single logical qubit per code block, while extension to $k>1$ is conceptually possible. 

For a single-qubit target operation $U$, we call an ansatz for the physical realization $U_L$ \emph{strictly-transversal} if it is implemented by the same single-qubit unitary $V$ on all $n$ physical qubits, i.e.,
\begin{align}
    \label{eq:strictly_transversal_1q}
    U_L = V^{\otimes n}, ~\text{with } V \in \mathrm{U}(2).
\end{align}
Relaxing this assumption by allowing the tensor product of potentially different single-qubit unitaries, we define a \emph{transversal} ansatz as
\begin{align}
    \label{eq:transversal_1q}
    U_L = \bigotimes_{i}^{n} V_i, ~\text{with } V_i \in \mathrm{U}(2).
\end{align}
In the variational setting, these notions are enforced simply by constraining the ansatz $U_L(\Psi)$ as shown in \cref{subfig:ansatz_one_qubit}. Both variants lead to depth-one realizations on the code block and enjoy standard transversality-based fault-tolerance guarantees~\cite{eastin2009restrictions}. Our framework also supports more expressive \emph{non-transversal} ansätze for single-logical-qubit gates, where $U_L(\Psi)$ contains controlled two-qubit gates at randomized locations (similar to the \gls{rea} ansatz, see~\cite{meyer2025learning}). Such ans\"atze do no longer provide concrete fault tolerance guarantees, but are useful to identify low-depth realizations for \gls{eftqc}.

For a controlled two-logical-qubit operation $CU_L$ between a control block
and a target block of the same code, we consider the following hierarchy: If each physical control qubit interacts only with the corresponding target qubit via the same controlled gate, we again use the notion \emph{strictly-transversal}. Concretely, this reads as
\begin{align}
    \label{eq:strictly_transversal_2q}
    CU_L = (CV)^{\otimes n}, ~\text{with } V \in \mathrm{U}(2),
\end{align}
where $CV$ denotes the fixed controlled two-qubit unitary. Accordingly, we say the ansatz is \emph{transversal} if
\begin{align}
    \label{eq:transversal_2q}
    CU_L = \bigotimes_{i}^{n} CV_{i}, ~\text{with } V_i \in \mathrm{U}(2),
\end{align}
where the controlled two-qubit unitaries are potentially different. Lastly, we use the term \emph{weakly-transversal} if each control-target pair is acted on by an arbitrary two-qubit gate as
\begin{align}
    \label{eq:weakly_transversal_2q}
    CU_L = \bigotimes_{i}^{n} W_{i}, ~\text{with } W_i \in \mathrm{U}(4).
\end{align}
The weakly transversal case retains a strictly local, pairwise structure across the two blocks, but is of higher depth than the other two formulations. Concretely, we parameterize a generic two-qubit unitary $W_i$ using the standard decomposition $(V_{1} \otimes V_{2}) CZ (V_{3} \otimes V_{4}) CZ (V_{5} \otimes V_{6}) CZ (V_{7} \otimes V_{8})$, with $V_1, \dots, V_8 \in \mathrm{U}(2)$, and $CZ$ a controlled Pauli-$Z$ operation. This realization is still constant-depth and induces standard Eastin-Knill-style guarantees on fault tolerance~\cite{eastin2009restrictions}, but can impact practical thresholds negatively compared to the other transversality notions~\cite{aliferis2005quantum,knill2005quantum}.

\medskip\noindent
In summary, our learning framework can be instantiated with any ansatz class for $U_L(\Psi)$, from fully unconstrained multi-layer circuits to strictly transversal depth-one circuits. In practice, we often fix the ansatz to one of the transversal or weakly transversal forms above. This yields implementations that are not only easier to optimize and evaluate but also exhibit structural properties that carry fault tolerance guarantees for \gls{ftqc} and are also desirable for \gls{eftqc}.


\section{\label{sec:empirical}Numerical Validation on Stabilizer Codes}

We now concretely instantiate the pipeline from \cref{sec:method} on standard stabilizer codes with a single logical qubit per block. We fix known encoders $U_{\mathrm{enc}}$ of additive codes, train physical realizations of logical gates, and use known realizations from the literature as a reference.
Our goals in this section are to verify that the procedure can (re)discover known logical operations from this reference set, study the impact of the loss-function variants from \cref{subsec:method_loss} on optimization, and assess the expressivity of (strictly-)transversal ansätze, including overparameterization via layer repetition.
In \cref{sec:vareftqc}, we extend this pipeline by joint optimization of non-additive encodings, and also experiment with weakly-transversal and non-transversal circuit ans\"atze.

\subsection{\label{subsec:empirical_setup}Experimental Setup}

\begin{table}[tb]
    \caption{\label{tab:stabilizer_transversal}Standard stabilizer codes with single logical qubits considered in this work. For each code and each target intra-block gate $X_L,Z_L,H_L,S_L,T_L$ and inter-block gate $CX_L,CZ_L,CS_L$, a double check mark \cmark\cmark~indicates that a strictly-transversal realization of the respective logical gate exists, while a single check mark \cmark~indicates the existence of a transversal realization. This data represents the ground truth extracted from the literature; the corresponding results obtained with our learning pipeline are shown in \cref{fig:operation_method} and agree perfectly with this ground truth. In particular, the procedure rediscovers all existing transversal realizations, apart from the two-qubit gates for the $[[15,1,3]]$ code, which we excluded due to computational complexity.}
    \begin{ruledtabular}
        \begin{tabular}{c|cccccccc}
            \multirow{2}{*}{\textbf{Code}} & \multicolumn{8}{c}{\textbf{Operation}} \\
            & ~$X_L$~ & ~$Z_L$~ & ~$H_L$~ & ~$S_L$~ & ~$T_L$~ & $CX_L$ & $CZ_L$ & $CS_L$ \\
            \hline
            $[[3,1,1]]$ & \cmark\cmark & \cmark\cmark & $-$ & \cmark & \cmark & \cmark\cmark & \cmark\cmark & \cmark \\
            $[[4,1,2]]$ & \cmark & \cmark & $-$ & $-$ & $-$ & \cmark\cmark & $-$ & $-$ \\
            $[[5,1,3]]$ & \cmark\cmark & \cmark\cmark & $-$ & $-$ & $-$ & $-$ & $-$ & $-$ \\
            $[[7,1,3]]$ & \cmark\cmark & \cmark\cmark & \cmark\cmark & \cmark & $-$ & \cmark\cmark & \cmark\cmark & $-$ \\
            $[[9,1,3]]$ & \cmark & \cmark & $-$ & $-$ & $-$ & ~\cmark\cmark\footnotemark[1] & $-$ & $-$ \\
            $[[10,1,4]]$ & \cmark & \cmark & $-$ & $-$ & $-$ & $-$ & $-$ & $-$ \\
            $[[11,1,5]]$ & \cmark & \cmark & $-$ & $-$ & $-$ & $-$ & $-$ & $-$ \\
            $[[15,1,3]]$ & \cmark\cmark & \cmark\cmark & $-$ & \cmark & \cmark & {\color{gray}\cmark\cmark} & {\color{gray}\cmark} & {\color{gray}\cmark} \\
        \end{tabular}
    \end{ruledtabular}
    \footnotetext[1]{Control and target order inverted for physical realization.}
\end{table}

We consider the stabilizer codes introduced in \cref{subsec:preliminaries_qec} and also listed in \cref{tab:stabilizer_transversal}. For each code $\mathcal{C}$ we only assume access to its encoding unitary $U_{\mathrm{enc}}$, i.e., no stabilizer description, decoder, or recovery circuit is used. The strictly-transversal and transversal logical gates known from the literature for these codes are also compiled in \cref{tab:stabilizer_transversal} and define the instances we study.

We train the intra-block single-qubit logical gates $X_L, Z_L, H_L, S_L, T_L$ and inter-block two-qubit logical gates $CX_L, CZ_L, CS_L$ using the pipeline described before: logical input states are sampled from the single- and two-qubit 2-designs, and we optimize the respective loss variant following \cref{eq:training_minimization_pariwise} using a quasi-Newton L-BFGS optimizer~\cite{liu1989limited} with a history size of $100$, and $20$ internal iterations per epoch. The one-qubit factors $V$ appearing in \cref{eq:strictly_transversal_1q,eq:transversal_1q,eq:strictly_transversal_2q,eq:transversal_2q,eq:weakly_transversal_2q} are parameterized by Euler-angle $U3(\theta,\phi,\lambda)$ gates. The implementation uses mainly the \texttt{PyTorch} framework~\cite{paszke2019pytorch} for the training procedure itself, and \texttt{PennyLane}~\cite{bergholm2018pennylane} and the \texttt{qiskit-torch-module}~\cite{meyer2024qiskit} for the quantum subroutines.

As a quality measure for evaluation, we use the worst-case 2-design loss $\overline{\mathcal{O}}_{\mathcal{S}}(C;\Psi)$: a run is declared \emph{successful} if $\overline{\mathcal{O}}_{\mathcal{S}}(C;\Psi) \leq 10^{-5}$. For each configuration (code, gate, loss variant) we perform $100$ independent runs with random initialization and report the \emph{success percentage}, i.e., the fraction of runs that satisfy the threshold.
For the loss comparison in \cref{fig:operation_method}, we aggregate over all instances from \cref{tab:stabilizer_transversal} that are numerically feasible: $22$ one-qubit and $7$ two-qubit code–gate combinations. Three additional two-qubit instances for the $[[15,1,3]]$ code are too costly to include in this sweep, as this would require simulating $30$-qubit systems.

\subsection{\label{subsec:empirical_loss}Impact of Loss Function and Overparameterization}

\begin{figure}[tb]
    \centering
    \subfigure[\label{subfig:operation_loss}Impact of loss function variant on success percentage.]{
        \includegraphics{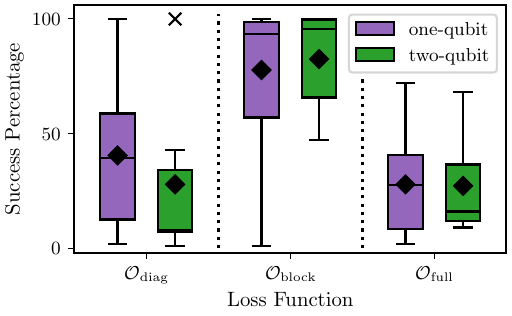}
    }\\
    \subfigure[\label{subfig:operation_repeat}Impact of ansatz overparameterization on success percentage.]{
        \includegraphics{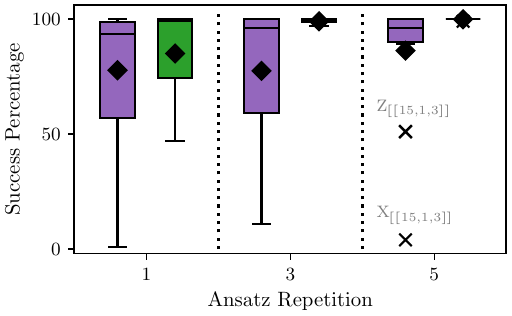}
    }
    \caption{\label{fig:operation_method}Numerical success rates for learning transversal realizations of logical operations on stabilizer codes. Evaluations are performed on the code-operation pairs listed in \cref{tab:stabilizer_transversal}, separated into one-qubit intra-block and two-qubit inter-block gates. Both subfigures show histograms of success percentages for the $22$ one-qubit and $7$ two-qubit setups; the black horizontal line indicates the median, while the diamond represents the average. In (a), we compare training with the three loss variants $\mathcal{O}_{\mathrm{diag}}$, $\mathcal{O}_{\mathrm{block}}$, and $\mathcal{O}_{\mathrm{full}}$, revealing that the block-diagonal formulation is clearly superior to the other two. In (b), we further analyze $\mathcal{O}_{\mathrm{block}}$ for overparameterized transversal ans\"atze, i.e., multi-layer repetitions of the parameterized single- and two-qubit gates. While some outliers remain, in particular the $X_L$ gate in the $[[15,1,3]]$ code, which converged only for a single run, the performance improvement saturates at around $3$ to $5$ repetitions. The exact success counts for each setup are reported in \cref{app:empirical}.}
\end{figure}

We first validate the pipeline against the instances listed in \cref{tab:stabilizer_transversal}, where the existence of strictly-transversal and transversal realizations is extracted from literature and confirmed with existing tools~\cite{webster2023transversal}. Across all code-operation pairs, this ground truth is consistent with our findings, i.e., we rediscover transversal realizations where they exist, and confirm the non-existence of such realizations in the other cases. This exact agreement, i.e., recovering all existing transversal realizations and not hallucinating any operations that should not exist, provides a strong consistency check for our proposed procedure.

Next, we analyze the choice of loss function variant that is used for training, with the three versions $\mathcal{O}_{\mathrm{diag}}$, $\mathcal{O}_{\mathrm{block}}$, and $\mathcal{O}_{\mathrm{full}}$, as introduced in \cref{subsec:method_loss}. The results are reported as histograms of success percentages as defined above in \cref{subfig:operation_loss}. We find that the choice of loss function has a pronounced effect on optimization. In \cref{app:loss} we argue that this is mostly caused by bad local optima in the loss landscape. While moving to the full-pair loss $\mathcal{O}_{\mathrm{full}}$ removes some degeneracies, in practice the improvement is limited or even non-existent, likely due to increased complexity and largely redundant information. In contrast, the block-diagonal loss $\mathcal{O}_{\mathrm{block}}$, which augments $\mathcal{O}_{\mathrm{diag}}$ with pairs of mutually orthogonal states, clearly outperforms both alternatives in terms of median and average success rate for both one- and two-qubit gates, and also reduces variability. We therefore adopt $\mathcal{O}_{\mathrm{block}}$ as the default loss in the remainder of this work.

To further improve trainability without relaxing transversality, we also consider overparameterized transversal ansätze obtained by repeating the same layer $r$ times with independent parameters, i.e., 
\begin{align}
    \label{eq:transversal_1q_over}
    U_L = \bigotimes_{i}^{n} V_{i}^{(r)} \cdots V_{i}^{(1)}, ~\text{with } V_{i}^{(r)} \in \mathrm{U}(2).
\end{align}
Importantly, for the instantiation with universal single-qubit rotations used in this work, after training all $3 \cdot r$ parameters can, by construction, be condensed back to $3$ parameters of a single parameterized $U3$ gate. The same holds for controlled-$V$ gates in a transversal (but not weakly-transversal) two-qubit ansatz, i.e., the depth of the transversal operation at deployment time does not increase due to this adaptation. We analyze the success rate for the setups previously considered with $\mathcal{O}_{\mathrm{block}}$ for increasing repetition rates in \cref{subfig:operation_repeat}. Increasing $r$ substantially boosts success percentages, with the improvement saturating around a repetition rate of $3$ to $5$.
Some outliers remain; for example, the logical $X_L$ gate on $[[15,1,3]]$ converges only in a single run for $r=5$. This behavior is reminiscent of the effect of overparameterization in classical machine learning, where highly overparameterized models can be easier to optimize~\cite{du2019gradient,allen2019convergence}, but also might lead to \emph{double descent} behavior~\cite{hastie2022surprises}. Aggregated over all codes and gates, moderate ansatz repetition nonetheless can significantly improve success robustness while preserving transversal structure.

\bigskip\noindent
In summary, the experiments in this section demonstrate that our pipeline reliably rediscovers all known (strictly-)transversal logical operations on standard stabilizer codes. Among the considered loss variants, the block-diagonal loss $\mathcal{O}_{\mathrm{block}}$ yields the most stable and accurate training behavior, and moderate overparameterization via ansatz repetition further improves success rates while preserving transversal structure. The full numerical success statistics underlying the reported results are provided in \cref{app:empirical}. In the next section, \cref{sec:vareftqc}, we build on these insights to jointly optimize non-additive encodings and logical operations.


\section{\label{sec:vareftqc}Variational Early Fault-Tolerant Quantum Computing}

The previous sections established the variational pipeline for learning logical operations on arbitrary codes, given access only to the encoding unitary. At the same time, the \gls{varqec} procedure~\cite{meyer2025learning,meyer2025variational} for tailoring encodings to noise structures has been summarized in \cref{sec:preliminaries}. Now, we combine both ingredients into a single co-design procedure that simultaneously learns a noise-tailored non-additive code and a set of logical gates. We refer to this combined approach as \emph{\glsentrylong{vareftqc}} (VarEFTQC). The term emphasizes that we target small, noise-tailored, typically non-additive codes $((n,2))$ with transversal or at least shallow-depth logical realizations, in the spirit of the \gls{eftqc} paradigm~\cite{katabarwa2024early}. Our goal is not to perform a full threshold analysis or to construct a universal, strictly fault-tolerant gate set for asymptotically large architectures, but rather to identify practical logical encodings and gate gadgets that significantly mitigate realistic noise at modest overhead.

Algorithmically, \gls{vareftqc} is realized as a single training loop that optimizes the encoder, as in \gls{varqec}, and establishes physical realizations for a chosen set of logical gates. The full pipeline is implemented in a \texttt{VarEFTQC} software framework; details are provided in the data availability statement.


\subsection{\label{subsec:vareftqc_joint}Joint Optimization of Encodings and Operations}

Let $\mathcal{N}$ be a fixed physical noise channel. Using \gls{varqec}, a parameterized encoder $U_{\mathrm{enc}}(\Theta)$ defines a code $\mathcal{C}_\Theta$ by mapping $k$ logical qubits and $n-k$ ancillas to an $n$-qubit codespace, in a way that minimizes the worst-case distinguishability loss $\overline{\mathcal{D}}(\mathcal{N};\Theta)$ via optimizing the average-case 2-design proxy $\mathcal{D}_{\mathcal{S}}(\mathcal{N};\Theta)$ (for details see \cref{subsec:preliminaries_VarQEC} and Refs.~\cite{meyer2025learning,meyer2025variational}). For a given encoding $\mathcal{C}_\Theta$ and a target unitary $U$, \cref{sec:method} introduced the worst-case operation loss $\overline{\mathcal{O}}^U(\mathcal{C}_\Theta;\Psi^U)$, which is most efficiently minimized via the average-case block-diagonal proxy $\mathcal{O}^U_{\mathrm{block}}(\mathcal{C}_{\Theta};\Psi^U)$, as analyzed in \cref{sec:empirical}. If the worst-case loss is below a threshold of $\varepsilon = 10^{-5}$, we say that the physical realization $U_L(\Psi^U)$ implements the target $U$ within the codespace.

To jointly tailor the encoding to~$\mathcal{N}$ and to learn a set of logical operations $\{U\}$, we define the regularized composite loss
\begin{align}
    \begin{split}
        \label{eq:vareftqc_general}
        \mathcal{V}(\mathcal{N};\Theta,\Psi) =& \mathcal{D}_{\mathcal{S}}(\mathcal{N};\Theta) \\
        &+ \gamma \sum_{U} \mathcal{O}^{U}_{\mathrm{block}}(\mathcal{C}_{\Theta};\Psi^{U}),
    \end{split}
\end{align}
where $\Psi = \{\Psi^U\}_U$ collects all parameters of the operation ans\"atze. The code $\mathcal{C}_\Theta$ in \cref{eq:vareftqc_general} is thus \emph{not} fixed during training: changing~$\Theta$ modifies both the encoding and the logical subspace on which all operation losses are evaluated. The regularization parameter $\gamma \geq 0$ controls the relative importance between noise-tailoring and gate synthesis. For small $\gamma$, the optimization primarily reproduces the \gls{varqec} behavior and focuses on preserving logical distinguishability under $\mathcal{N}$, whereas larger $\gamma$ increasingly encourages the existence of high-fidelity logical implementations of the chosen gates. Conceptually, one may schedule~$\gamma$ during training, starting with a small value to discover a rough noise-adapted code and gradually increasing it to enforce that the logical operations reach a target quality threshold~$\varepsilon$. For the proof-of-concept experiments in this work, we identified a static $\gamma = \frac{1}{2 \cdot \left| \lbrace U \rbrace \right|}$ to be sufficient for stable convergence.

In practice, we found another technique to have a big impact on the convergence speed of \gls{vareftqc}: Instead of starting training from all-random parameter sets $\Theta,\Psi$, we pre-train the encoder $U_{\mathrm{enc}}(\Theta)$ on $\mathcal{N}$ using pure \gls{varqec}, yielding a noise-tailored encoding with parameters $\Theta^{\star}$, whose associated code $\mathcal{C}_{\Theta^{\star}}$ does not necessarily support the target logical operations. Subsequently, we use these pre-trained parameters as warm-start initializations~\cite{meyer2024warm} for the full \gls{vareftqc} training routine in \cref{eq:vareftqc_general}. The parameters $\Theta^{\star}$ are then re-trained (one could also interpret this as finetuning) together with $\Psi$ during this procedure. Across all noise models and gate sets considered in this work, we observe a reduction in training steps when using this warm-start approach over the baseline cold-start initialization, also when accounting for the initial overhead of pre-training the encoding.


\subsection{\label{subsec:vareftqc_iqp}Noise-Tailored Codes with Native IQP Support}

We first apply \gls{vareftqc} to construct non-additive codes with native support for logical \gls{iqp}-type circuits. \Gls{iqp} circuits consist of an initial layer of Hadamard gates, followed by a polynomial number of commuting diagonal gates in the computational basis, and finally a last layer of Hadamards before measurement~\cite{nest2009simulating,bremner2011classical}. More concretely, this can be expressed as
\begin{align}
    \label{eq:iqp}
    H^{\otimes n} \prod_{i}^{k} d_i H^{\otimes n} \ket{0}^{\otimes n},
\end{align}
where $k$ is polynomial in $n$ and all gates $d_i$ are $Z$-diagonal. Such circuits are believed to be classically hard to sample from under reasonable complexity-theoretical assumptions~\cite{bremner2011classical}. In particular, this already holds for \emph{sparse} \gls{iqp}-families with gates $d_i \in \lbrace T, CS \rbrace$, i.e., at most two-body interactions~\cite{bremner2017achieving}. Interestingly, certain expectation values can be estimated efficiently classically, constituting \gls{iqp}-type circuit families as promising candidates for near-term practical quantum advantage, in particular in the context of quantum generative models~\cite{coyle2020born,recio2025train}.

From the perspective of \gls{qec} and \gls{vareftqc}, \gls{iqp} circuits are particularly attractive because all non-Hadamard operations are diagonal in the computational basis and can, in principle, be realized transversally on suitable codes without violating the Eastin-Knill theorem~\cite{paletta2024robust,hangleiter2025fault}. In our setting, state preparation in a logical Hadamard state~$\ket{+}_L$ via \gls{varqec} provides the initial Hadamard layer, and the final Hadamard can be absorbed into a change of measurement basis. The core requirement is thus to realize logical $Z$-diagonal gates such as $T_L$, $CZ_L$, and $CS_L$. Concretely, we consider the following \gls{iqp}-type logical gate sets:
\begin{align}
    &\lbrace CZ_L, T_L \rbrace \label{eq:iqp_cz}\\
    &\lbrace CS_L, T_L \rbrace \label{eq:iqp_cs}\\
    &\lbrace CX_L, T_L \rbrace \label{eq:iqp_cx}
\end{align}
Hereby, we aim to realize $CZ_L$, $CS_L$, and $CX_L$, i.e., logical controlled-$Z$, controlled-$S$, and controlled-$X$ gates, as inter-block operations. Hereby, \cref{eq:iqp_cz} is the most natural $Z$-diagonal choice; however, to the best of our knowledge, no sampling hardness results have been established for this circuit type. To that effect, the finer granularity in \cref{eq:iqp_cs}, with $CS = \sqrt{CZ}$, guarantees hardness of classical sampling under some additional assumptions on the actual gate placements~\cite{bremner2017achieving}. Lastly, \cref{eq:iqp_cx} can be used to realize the same circuit family, employing the identity $CS = CX (T \otimes T^{\dagger})  CX (I \otimes T) $~\cite{paletta2024robust}.

\begin{figure}[htb!]
    \centering
    \subfigure[\label{subfig:training_curve_encoding}Training encoding on $\mathcal{N}_{\mathrm{adep}}$ noise using \gls{varqec}.]{
        \includegraphics{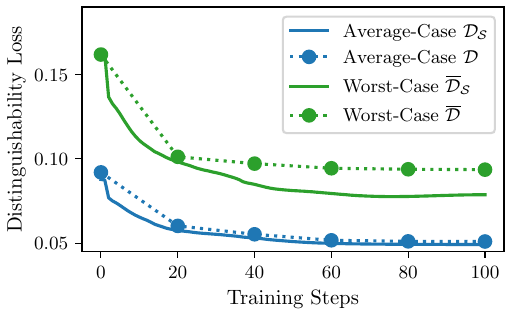}
    }\\
        \subfigure[\label{subfig:training_curve_operation}Training encoding and operations using warm-start \gls{vareftqc}.]{
        \includegraphics{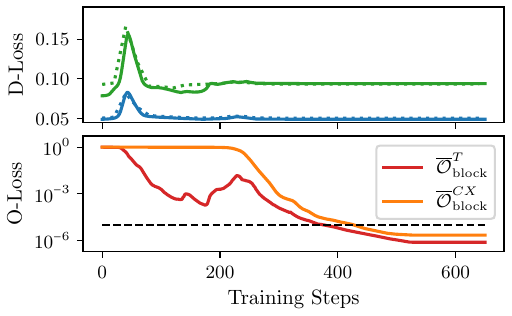}
    }\\
    \subfigure[\label{subfig:training_curve_encoding_operation}Training encoding and operations using \gls{vareftqc}.]{
        \includegraphics{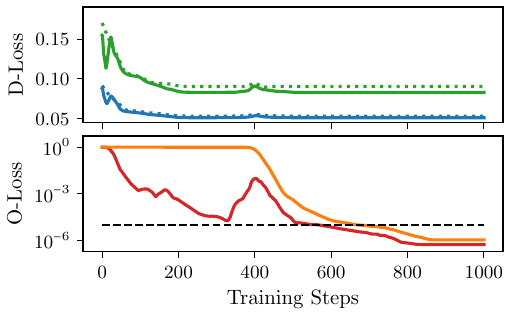}
    }
    \caption{\label{fig:training_curve}
        Training behaviour of \gls{vareftqc} for a $((5,2))$ code under asymmetric depolarizing noise $\mathcal{N}_{\mathrm{adep}}$ with noise strength $p=0.1$, asymmetry $c=0.5$, and target logical gate set $\{T_L, CX_L\}$. The ansatz for $T_L$ is transversal, while the ansatz for $CX_L$ is weakly transversal.  
        (a) Encoding-only optimization using \gls{varqec}, with the average-case 2-design proxy $\mathcal{D}_{\mathcal{S}}$ as training objective; worst-case and validation (dashed) curves are shown in addition. (b) Warm-start \gls{vareftqc}: the parameters of the noise-tailored encoder from (a) initialize the encoding, and the average-case block-diagonal operation loss $\mathcal{O}_{\mathrm{block}}$ is added with regularization weight $\gamma=\frac{1}{4}$.
        (c) Same \gls{vareftqc} setup as in (b), but without warm-start initialization of the encoding.
    }
\end{figure}

\begin{table}[tb]
    \caption{\label{tab:gateset_iqp}
        Logical IQP-type gate sets obtained with \gls{vareftqc}. Non-additive codes $((n,2))$ with varying number of physical qubits are tailored to $\mathcal{N}_{\mathrm{adep}}$ and $\mathcal{N}_{\mathrm{bit}}$, while simultaneously one of the gate sets $\lbrace CZ_L, T_L \rbrace$, $\lbrace CS_L, T_L \rbrace$, and $\lbrace CX_L, T_L \rbrace$ is learned within the codespace. With $\overline{\mathcal{D}}_{\downarrow}$, we denote the relative reduction of information loss under encoding, i.e.\ the worst-case distinguishability loss for the pure physical qubit under the noise channel, divided by the respective loss for the encoded logical system.        
        A double check mark \cmark\cmark~indicates that a strictly-transversal ansatz was used, a single check mark \cmark~indicates the use of a transversal ansatz, and brackets (\cmark) refer to a weakly-transversal ansatz; for some instances, we denote two alternative combinations of ansatz structures. Training curves for tailoring an encoding to $\mathcal{N}_{\mathrm{adep}}$ noise with joint construction of the gate set $\lbrace CX_L, T_L \rbrace$ are shown in \cref{fig:training_curve}.
    }
    \begin{ruledtabular}
        \begin{tabular}{ccc|cc|cc|cc}
            \multirow{2}{*}{\textbf{Noise}} & \multirow{2}{*}{\textbf{Code}} & \multirow{2}{*}{$\overline{\mathcal{D}}_{\downarrow}$} & \multicolumn{6}{c}{\textbf{IQP Gate Set}} \\
            & & & \multicolumn{2}{c|}{$CZ_L$~\&~$T_L$~~} & \multicolumn{2}{c|}{$CS_L$~\&~$T_L$~~} & \multicolumn{2}{c}{$CX_L$~\&~$T_L$~~} \\
            \hline
            \multirow{3}{*}{$\mathcal{N}_{\mathrm{adep}}$} & \multirow{1}{*}{$((4,2))$} & $1.9$ & (\cmark) & \cmark & (\cmark) & \cmark & (\cmark) & \cmark \\
            \cline{2-9}
            & \multirow{1}{*}{$((5,2))$} & $2.0$ & (\cmark) & \cmark & (\cmark) & \cmark & (\cmark) & \cmark \\
            \cline{2-9}
            & \multirow{1}{*}{$((6,2))$} & $2.4$ & (\cmark) & \cmark & (\cmark) & \cmark & (\cmark) & \cmark \\
            \hline
            \multirow{4}{*}{$\mathcal{N}_{\mathrm{bit}}$} & \multirow{2}{*}{$((3,2))$} & \multirow{2}{*}{$3.6$} & \cmark\cmark & \cmark & \multirow{2}{*}{\cmark} & \multirow{2}{*}{\cmark\cmark} & \cmark\cmark & \cmark \\
            & & & \cmark & \cmark\cmark & & & (\cmark) & \cmark\cmark \\
            \cline{2-9}
            & \multirow{2}{*}{$((5,2))$} & \multirow{2}{*}{$11.8$} & \multirow{2}{*}{\cmark\cmark} & \multirow{2}{*}{\cmark\cmark} & \cmark\cmark & \cmark & \cmark\cmark & \cmark \\
            & & & & & \cmark & \cmark\cmark & (\cmark) & \cmark\cmark \\
        \end{tabular}
    \end{ruledtabular}
\end{table}

We evaluate \gls{vareftqc} on two noise models: asymmetric depolarizing noise $\mathcal{N}_{\mathrm{adep}}$ with overall noise strength $p=0.1$ and asymmetry factor $c=0.5$~\cite{meyer2025learning,olle2024simultaneous}. In the form of \cref{eq:pauli_noise}, this boils down to $p_X = p_Y = 0.007$ and $p_Z = 0.086$. Moreover, bitflip noise $\mathcal{N}_{\mathrm{bit}}$ of strength $p=0.1$. Both noise models are applied independently to all physical qubits. The quality of an encoding is measured by the suppression of information loss compared to a bare physical execution. In particular, we quantify this by the worst-case distinguishability loss for the physical qubit, divided by the same measure for the encoded system, denoted by $\overline{\mathcal{D}}_{\downarrow}$ in the following.

\medskip\noindent
In \cref{fig:training_curve}, we illustrate the training dynamics of \gls{vareftqc} for an $((5,2))$ code under $\mathcal{N}_{\mathrm{adep}}$ with target gate set $\{T_L, C X_L\}$. The ansatz for $T_L$ is transversal with three repetitions following \cref{eq:transversal_1q_over}, while $CX_L$ is realized by a weakly-transversal ansatz. \Cref{subfig:training_curve_encoding} shows pure \gls{varqec} pre-training of the encoding, by minimizing $\mathcal{D}_{\mathcal{S}}(\mathcal{N}_{\mathrm{adep}};\Theta)$ for $5$ epochs with $20$ L-BFGS steps each. As a baseline, a bare physical qubit obtains $\overline{\mathcal{D}}(\mathcal{N}_{\mathrm{adep}}) \approx 0.185$, while the noise-tailored $5$-qubit encoding achieves $\overline{\mathcal{D}}(\mathcal{N}_{\mathrm{adep}};\Theta^{\star}) \approx 0.091$, resulting in $\overline{\mathcal{D}}_{\downarrow} \approx 2.03$. For comparison, the $[[5,1,3]]$ perfect code achieves only $\overline{\mathcal{D}}_{\downarrow} \approx 1.35$ on this noise setup.

In \cref{subfig:training_curve_operation}, we perform warm-start \gls{vareftqc} by initializing $\Theta \leftarrow \Theta^\star$ from the pre-training and optimize the joint objective
\begin{align}
        &\mathcal{D}_{\mathcal{S}}(\mathcal{N};\Theta) + \frac{1}{4} \left(  \mathcal{O}^{T}_{\mathrm{block}}(\mathcal{C}_{\Theta};\Psi^{T}) + \mathcal{O}^{CX}_{\mathrm{block}}(\mathcal{C}_{\Theta};\Psi^{CX}) \right)
\end{align}
for $32$ epochs. While the distinguishability loss of the encoding initially increases again, it quickly converges to essentially the same level as encoding-only training around $\overline{\mathcal{D}}_{\downarrow} \approx 2.0$. Both operation loss values converge below the targeted threshold of $10^{-5}$. The training curves in \cref{subfig:training_curve_encoding_operation} denote $50$ epochs of \gls{vareftqc} without warm-start initialization, reaching comparable final values, but requiring significantly more training steps to do so. Accounting for pre-training, the convergence time by using warm-start \gls{vareftqc} was observed to be reduced by about $30-50\%$ across multiple instances. Overall, this example demonstrates that the joint loss $\mathcal{V}(\mathcal{N};\Theta,\Psi)$ can simultaneously identify a noise-tailored encoding and high-fidelity logical gates, without sacrificing the error suppression capabilities of the code.

In \cref{tab:gateset_iqp}, we summarize the \gls{vareftqc} results for \gls{iqp}-type gate sets across different code sizes and both noise models. For each configuration, we report the information-loss reduction $\overline{\mathcal{D}}_\downarrow$ and the most compact transversality structure in which the gate set could be realized. In the case of $\mathcal{N}_{\mathrm{adep}}$ noise, we learn codes with $n=4$ to $6$ physical qubits and consider native, i.e., transversal, realizations of the three \gls{iqp}-type gate sets. Interestingly, in all cases, the two-qubit gate could only be realized weakly-transversally, while the single-qubit $T_L$ gate allows for a transversal implementation. This invariance suggests that the dominant constraints are dictated by the noise structure, and not necessarily the code parameters. However, there likely are \emph{transition points} that enable the construction of certain transversality combinations, such as the minimal number of $n=15$ qubits to realize $T_L$ transversally in stabilizer \gls{qec} codes~\cite{koutsioumpas2022smallest}. This can actually be observed for $\mathcal{N}_{\mathrm{bit}}$ noise, where the $5$-qubit version allows for a wider range of transversality combinations on two of the three gate sets compared to the $3$-qubit version, additionally to a significantly higher noise suppression.

In summary, \gls{vareftqc} successfully discovers families of small non-additive codes that are adapted to realistic noise models and admit native logical implementations of \gls{iqp}-type gate sets with transversal structure. This makes the procedure an attractive candidate for \gls{eftqc} realizations of classically hard sampling tasks and quantum generative models.


\subsection{\label{subsec:vareftqc_universal}Towards Universal Logical Gate Sets}

\begin{table}[tb]
    \caption{\label{tab:gateset_universal}
        Logical universal gate sets obtained with \gls{vareftqc}. Non-additive codes $((5,2))$ are tailored to $\mathcal{N}_{\mathrm{adep}}$ and $\mathcal{N}_{\mathrm{bit}}$, while simultaneously the gate set $\lbrace CX_L, T_L, H_L \rbrace$ is learned within the codespace. With $\overline{\mathcal{D}}_{\downarrow}$, we denote the relative reduction of information loss under encoding, i.e.\ the worst-case distinguishability loss for the pure physical qubit under the noise channel, divided by the respective loss for the encoded logical system.        
        A single check mark \cmark~indicates the use of a transversal ansatz, brackets (\cmark) refer to a weakly-transversal ansatz, and a cross \xmark~denotes a non-transversal ansatz, with the subscript quantifying the number of two-qubit blocks violating transversality (see \cref{subfig:ansatz_one_qubit}); for some instances, we denote multiple alternative combinations of ansatz structures.
    }
    \begin{ruledtabular}
        \begin{tabular}{ccc|cccccc}
            \multirow{2}{*}{\textbf{Noise}} & \multirow{2}{*}{\textbf{Code}} & \multirow{2}{*}{$\overline{\mathcal{D}}_{\downarrow}$} & \multicolumn{5}{c}{\textbf{Universal Gate Set}} \\
            & & & ~$CX_L$~ & \& & ~~$T_L$~~ & \& & ~~$H_L$~~ \\
            \hline
            $\mathcal{N}_{\mathrm{adep}}$ & $((5,2))$ & $2.0$ & (\cmark) & & \cmark & & ~\xmark$_1$ \\
            \hline
            \multirow{4}{*}{$\mathcal{N}_{\mathrm{bit}}$} & \multirow{4}{*}{$((5,2))$} & \multirow{4}{*}{$11.8$} & (\cmark) & & \cmark & & ~\xmark$_1$ \\
            & & & (\cmark) & & \cmark\cmark & & ~\xmark$_1$ \\
            & & & \cmark & & \cmark & & ~\xmark$_3$ \\
            & & & \cmark\cmark & & \cmark & & ~\xmark$_5$ \\
        \end{tabular}
    \end{ruledtabular}
\end{table}

Finally, we investigate whether \gls{vareftqc} can be pushed from \gls{iqp}-type gate sets towards \emph{universal} logical gate sets, thereby moving closer to a fully-fledged \gls{eftqc} architecture. Concretely, we target the standard universal gate set
\begin{align}
  \{C X_L, T_L, H_L\},
\end{align}
which extends \cref{eq:iqp_cx} by a logical Hadamard $H_L$. The joint \gls{vareftqc} loss for this gate set, therefore, reads as
\begin{align}
    \begin{split}
        \mathcal{D}_{\mathcal{S}}(\mathcal{N};\Theta) + \frac{1}{6} \big( & \mathcal{O}^{H}_{\mathrm{block}}(\mathcal{C}_{\Theta};\Psi^{H}) \\ +&\mathcal{O}^{T}_{\mathrm{block}}(\mathcal{C}_{\Theta};\Psi^{T}) \\
        +&\mathcal{O}^{CX}_{\mathrm{block}}(\mathcal{C}_{\Theta};\Psi^{CX}) \big)
    \end{split}
\end{align}
We restrict the ans\"atze for $T_L$ and $CX_L$ to combinations that have been identified to be viable previously, see \cref{tab:gateset_iqp}. We again tailor the noise structures $\mathcal{N}_{\mathrm{adep}}$ and $\mathcal{N}_{\mathrm{bit}}$ from the previous subsection, and fix the code parameters to $((5,2))$.

We initially experimented with also restricting the ansatz for $H_L$ to be transversal, but no non-trivial code could be found that both realizes all three logical gates and suppresses the logical noise rate below the physical level. From a conceptual viewpoint, the failure to identify a fully transversal universal gate set on these \gls{vareftqc} codes is consistent with the Eastin-Knill no-go theorem and its approximate generalizations~\cite{faist2020continuous}. However, this should be interpreted as a purely empirical indication that approximate Eastin-Knill restrictions on universal transversal gate sets apply, as formally, we did not prove or analyze code distances. 

To proceed, we investigate a controllable violation of transversality in the $H_L$ realization, in particular, a small number of two-qubit gates as sketched in \cref{subfig:ansatz_one_qubit}. We summarize our findings in \cref{tab:gateset_universal}, again tailoring the noise structures $\mathcal{N}_{\mathrm{adep}}$ and $\mathcal{N}_{\mathrm{bit}}$ from the previous subsection. The best solutions found by \gls{vareftqc} that do not impact $\overline{\mathcal{D}}_{\downarrow}$ require a non-transversal ansatz with a
bounded number of two-qubit blocks. A single violating block is sufficient for $\mathcal{N}_{\mathrm{adep}}$, where $CX_L$ is realized weakly-transversally, and $T_L$ transversally. For $\mathcal{N}_{\mathrm{bit}}$, the number of violating blocks depends on the concrete transversality instances: assuming $CX_L$ to be only weakly-transversal, and $T_L$ realized with arbitrary transversality structure, also only a single two-qubit block is sufficient to realize $H_L$. In cases where $CX_L$ is transversal or even strictly transversal, the number of required violations increases to three and five, respectively. Importantly, all non-transversal $H_L$ realizations remain shallow and only introduce limited interaction, with some balancing options regarding the transversality structure of the remaining gates. While not universal fault-tolerance in a strict sense, this is still a desirable property for \gls{eftqc}, where the goal is to reduce circuit depth and noise sensitivity~\cite{katabarwa2024early}. In particular, the combination of (weakly-)transversal $CX_L$ and $T_L$ with a low-depth non-transversal $H_L$ can still lead to substantial improvements over unencoded circuits for realistic noise strengths.

\bigskip\noindent
In summary, this section introduced \gls{vareftqc} as a co-design framework that jointly optimizes noise-tailored (typically non-additive) encodings and their logical gate sets by combining the \gls{varqec} distinguishability objective with the block-diagonal operation loss in a single composite loss. Warm-starting from VarQEC-trained encoders preserves the noise-suppression of encoding-only training while improving convergence. The framework enables learning transversal realizations for IQP-type and low-depth realizations for universal logical gate sets. 


\section{\label{sec:discussion}Discussion of Progress and Challenges Towards Early FTQC}

Early fault-tolerant quantum computing (EFTQC)~\cite{katabarwa2024early} aims to obtain concrete logical advantages at modest code sizes, non-negligible noise rates, and under tight hardware constraints, instead of relying on asymptotic threshold guarantees. Small logical patches, shallow gadgets, and noise-aware design are central in this setting. In this section, we discuss how the results of \cref{sec:method,sec:empirical,sec:vareftqc} contribute to this agenda, and which open challenges remain.


\subsection{\label{subsec:discussion_codesign}Co-Designing Codes and Logical Gates for Early Fault-Tolerance}

A central ingredient for \gls{eftqc} is the ability to encode logical information in a way that is adapted to the actual noise processes of a given device. Realistic platforms rarely suffer from uniform depolarizing noise; instead, they exhibit biased, anisotropic, or otherwise structured errors~\cite{aliferis2008fault,tuckett2018ultrahigh}. Generic distance-$d$ codes are designed to correct arbitrary errors up to a certain weight and may therefore spend substantial overhead protecting against error types that almost never occur in practice. If the dominant noise modes are known, one would instead like to concentrate redundancy on those modes and trade absolute distance for lower qubit counts, shallower circuits, and better alignment with hardware constraints. The \gls{varqec} framework~\cite{meyer2025learning,meyer2025variational} realizes this idea by learning small encodings tailored to a given noise channel, via a distinguishability-based loss, and has produced $((n,K))$ codes that can match or surpass standard stabilizer codes at comparable $n$ for structured noise. However, \gls{varqec} on its own remains encoding-centric and does not specify any logical gate set.

The framework for learning logical operations developed in \cref{sec:method,sec:empirical} addresses this gap. Given only an encoding unitary $U_{\mathrm{enc}}$, it learns physical implementations of target logical unitaries by minimizing a fidelity-based operation loss over 2-designs, without requiring a stabilizer description, decoder, or explicit recovery circuit. The same formalism applies to intra-block and inter-block gates and can be instantiated with strictly-transversal, transversal, weakly-transversal, or shallow non-transversal ansätze. On standard stabilizer codes~\cite{gottesman1997stabilizer}, it rediscovers known transversal and strictly-transversal gates. Building on this, \gls{vareftqc} in \cref{sec:vareftqc} combines the \gls{varqec} distinguishability objective and the block-diagonal operation loss into a single co-design loop: instead of first learning a noise-tailored code and only then asking which gates it admits, it directly searches for encodings that both suppress information loss under the physical noise and admit high-fidelity, low-depth realizations of a prescribed logical gate set.

From a \gls{eftqc} perspective, the structural properties of these learned gates are as important as their fidelities. Transversal implementations are attractive because each physical fault can create at most one error per block~\cite{eastin2009restrictions}, so they inherit standard fault-tolerance properties and keep gadgets extremely shallow, whereas generic non-transversal circuits can spread a single fault into many errors within a block. Across the noise models considered, we observe a clear pattern: many single- and two-logical-qubit operations admit transversal realizations, while at least one gate needed for universality necessarily requires a non-transversal gadget. This behaviour is consistent with approximate Eastin--Knill constraints~\cite{eastin2009restrictions}, although we did not explicitly analyze the distance of the established codes. \Gls{vareftqc} effectively finds a favourable compromise by keeping the \emph{heavy-lifting} logical operations that appear most frequently in typical workloads transversal, and confining non-transversal structure to a small number of simple, low-depth gadgets. For \gls{eftqc}, this is the desirable regime: the bulk of logical operations is short, local, and structurally robust, while the few necessary non-transversal gadgets can be kept shallow.


\subsection{\label{subsec:discussion_challenges}Ways Forward and Open Challenges}

A practical limitation of the current implementation is that the training loop still relies on explicit classical simulation of encoded states and losses, which currently restricts direct optimization to small, classically tractable code patches. The following discussion therefore focuses on future extensions and open challenges that may partially mitigate this restriction by building larger logical architectures from such learned small components, rather than by directly training a single larger code block.

\textbf{Circuit-level evaluation.} A more complete assessment of the practical utility of the learned encodings and logical gadgets requires going beyond the isolated-gadget setting considered here and studying extended logical computations. This will require longer circuit-level simulations in which learned logical single- and two-qubit operations are interleaved with full \gls{qec} cycles, including syndrome measurement and recovery, under the relevant noise models. Such studies would more directly quantify the net logical benefit of the learned gadgets in extended computations and provide a clearer benchmark for comparing \gls{vareftqc} to more traditional code constructions. Carrying out this analysis is beyond the scope of the present work, but it is an important direction for future research.

\textbf{Code concatenation.} A central challenge is the further scalability of error suppression. Rather than naively increasing the size of a single trainable code patch, noise-aware concatenation~\cite{meyer2026learning} offers an alternate path: by estimating the effective logical noise after each level and tailoring the next-level code accordingly, we can reduce the error double-exponentially in the concatenation level~\cite{knill1998resilient,aliferis2005quantum}. In this setting, the interdependence of Ref.~\cite{meyer2026learning} with the present work is two-fold. On the one hand, small, noise-tailored codes with predominantly transversal or low-depth logical gates are natural base codes for the outer concatenation levels, where noise is still structured. On the other hand, realizing concatenation requires logical operations on learned codes: outer-level encoders, syndrome extraction routines, and logical gadgets must be implemented using logical gates of the inner codes. The \gls{vareftqc} framework provides exactly these ingredients.

\textbf{Patch-base logical qubits.} Independent of concatenation, our framework already supports scaling in the number of logical qubits via patch-based architectures. Each learned $((n,2))$ (or more generally $((n,K))$) code can be viewed as a logical patch, and \cref{sec:method,sec:empirical,sec:vareftqc} show that \gls{vareftqc} can learn controlled two-logical-qubit operations between patches, i.e., logical gates such as $\mathrm{CX}_L$ acting across two encoded blocks. This enables constructing multi-logical-qubit \gls{eftqc} architectures by wiring together small patches through learned inter-patch gates, in close analogy to how surface-code patches are coupled~\cite{fowler2012surface,horsman2012surface}, while keeping each patch small enough to be trainable with our 2-design-based losses. So far, however, we have assumed all-to-all connectivity within a patch and unconstrained coupling between patches. An important next step is to explicitly integrate realistic hardware layouts—such as grid structures for superconducting qubits, or shuttle-based connectivity in trapped-ion and neutral-atom platforms—into the \gls{vareftqc} ansätze. This would enable true patch-based logical architectures where both intra- and inter-patch gates are co-designed under locality constraints and compiled directly to the connectivity of a given device.

\textbf{Correlated Errors and Hardware Experiments.} A further engineering challenge is the extension of our framework to more complex, potentially correlated noise. Conceptually, neither the distinguishability loss nor the block-diagonal operation loss requires Pauli structure or independence; they apply to arbitrary \gls{cptp} maps. Results from \gls{varqec} already indicate that the same machinery extends natively to amplitude damping, thermal relaxation, correlated Pauli noise, and hardware-calibrated error models~\cite{meyer2025learning}. In practice, however, incorporating such channels into \gls{vareftqc} will be more expensive and may require refined 2-design proxies, Pauli-twirled approximations~\cite{geller2013efficient}, or sampling-based estimators to keep training tractable. Ultimately, it would be desirable to deploy \gls{vareftqc} codes on hardware. Doing so requires explicitly taking connectivity into account as discussed above, so that both encodings and logical operations are co-designed for the architecture under study.

\textbf{From Early to Full Fault-Tolerance.} Looking beyond \gls{eftqc} towards full-fledged \gls{ftqc}, a key conceptual gap is our current treatment of non-transversal gadgets. In this work, transversality serves as a structural proxy for fault tolerance, and shallow non-transversal gates are accepted as long as they remain short and local. For full \gls{ftqc}, however, it will be necessary to move beyond such proxies and to quantify fault tolerance of a given logical implementation more directly. A promising route is to augment the operation-learning objective with a measure of error propagation, so that candidate logical gates are penalized if faults inside the gadget spread too strongly within a code block. This would allow one to use the same optimization machinery to actively search for non-transversal logical operations that are not only accurate and low-depth, but also satisfy explicit fault-tolerance criteria, thereby interpolating between the \gls{eftqc}-oriented gadgets considered here and fully fault-tolerant logical architectures. From our perspective, the main challenge is to devise fault-tolerance measures that are both meaningful and efficiently evaluable.

\bigskip\noindent
Altogether, the co-design of encodings and logical gates, their integration with noise-aware concatenation, and the extensions to multi-patch architectures under connectivity constraints and realistic noise models sketched above outline a concrete roadmap from the present concept towards more scalable and reliable logical architectures. Systematically exploring this roadmap in detail is beyond the scope of this work and is left for future research.


\glsresetall
\section{\label{sec:summary}Summary and Conclusion}

In this paper, we have introduced a general variational framework for learning logical operations on arbitrary quantum error-correcting codes, assuming access only to the encoding unitary. Logical gates are represented by structured ansätze and trained via a fidelity-based \emph{operation loss} evaluated on 2-designs, without requiring a stabilizer description, decoder, or explicit recovery circuit. By formulating intra- and inter-block gates in a unified way, and by constraining the ansätze to transversal forms, the method naturally connects to fault-tolerance considerations while remaining applicable to non-additive codes.

On standard small stabilizer codes, the framework reliably rediscovers all known transversal logical gates. A comparison of loss variants shows that a pairwise block-diagonal formulation is crucial for stable training and high success rates, while moderate overparameterization of transversal ansätze further improves robustness without increasing logical circuit depth after compilation. Together, these results validate the basic learning pipeline and establish it as a proof-of-concept framework for synthesizing logical gates on existing codes.

Building on this, we combined a distinguishability-based objective for learning noise-tailored encodings~\cite{meyer2025learning} with the operation loss into a single co-design procedure. This concept, which we name \emph{\glsentrylong{vareftqc}} (VarEFTQC), simultaneously learns noise-tailored encodings and logical gate sets. For different biased noise models, VarEFTQC discovers small non-additive codes with native transversal support for IQP-type logical circuits, and identifies universal logical gate sets where only a single Hadamard gadget needs to violate transversality in a controlled, low-depth manner. 

In conclusion, our work highlights the importance of jointly tailoring encodings and logical operations to the underlying noise and hardware in advancing quantum error correction towards early fault-tolerant quantum computing. The present study should be viewed as a methods-driven proof of concept for this co-design paradigm in this early fault-tolerant regime, where resource efficiency and architectural adaptability are paramount. Future research can build upon this foundation to integrate VarEFTQC with noise-aware concatenation, patch-based logical architectures under restricted connectivity, and explicit fault-tolerance metrics to move from early to fully fault-tolerant quantum computing.


\section*{Data Availability Statement}
The introduced \texttt{VarEFTQC} pipeline is publicly available at \href{https://github.com/nicomeyer96/vareftqc}{https://github.com/nicomeyer96/vareftqc}. The repository also provides scripts to reproduce the plots in \cref{sec:vareftqc}. Learned VarEFTQC code instances, including encodings and logical gadgets, are available at \href{https://doi.org/10.5281/zenodo.20280560}{https://doi.org/10.5281/zenodo.20280560}. Further information and data are available upon reasonable request.

\begin{acknowledgments}
The authors gratefully acknowledge the scientific support and HPC resources provided by the Erlangen National High Performance Computing Center (NHR@FAU) of the Friedrich-Alexander-Universität Erlangen-Nürnberg (FAU). The hardware is funded by the German Research Foundation (DFG).

\smallskip
\noindent The authors used GPT-5.1 and GPT-5.4 for language editing of the paper. All content was reviewed and edited by the authors, who take full responsibility for the final work.

\smallskip
\noindent\textbf{Funding.} The research was supported by the German Federal Ministry of Research, Technology and Space, funding program Quantum Systems, via the project Q-GeneSys, grant number 13N17389. The research is also part of the Munich Quantum Valley (MQV), which is supported by the Bavarian state government with funds from the Hightech Agenda Bayern Plus.
\end{acknowledgments}


\appendix
\crefalias{section}{appsec}
\section{\label{app:geometric}Geometric Intuition of Loss Function and 2-design States}

\begin{figure}[tb]
    \centering
    \includegraphics{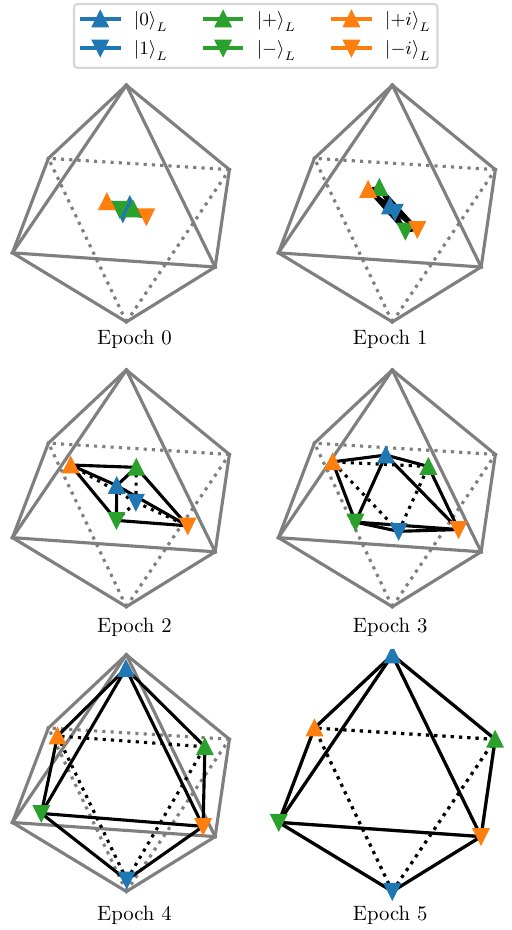}
    \caption{\label{fig:operation_loss_training}Illustration of training procedure on the 2-design octahedron. The prediction (black octahedron) is trained to fit the target logical operation on 2-design states (gray octahedron). In particular, the example shows the training of a logical $S$ operation on the $[[7,1,3]]$ code using the block-diagonal loss for $5$ epochs of $20$ L-BFGS iterations each.}
\end{figure}

\begin{figure}[tb]
    \centering
    \subfigure[\label{subfig:matrix_baseline_1q}Target gram matrix of pairwise fidelities.]{
        \includegraphics{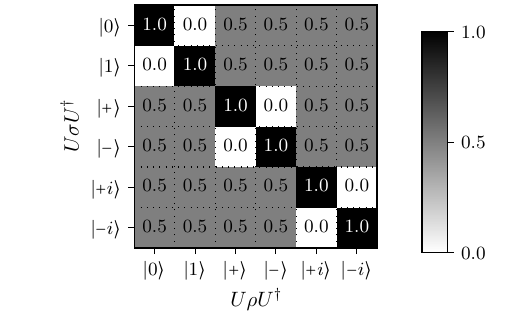}
    }\\
    \subfigure[\label{subfig:matrix_steane_S_1q}Sample prediction gram matrix of pairwise fidelities.]{
        \includegraphics{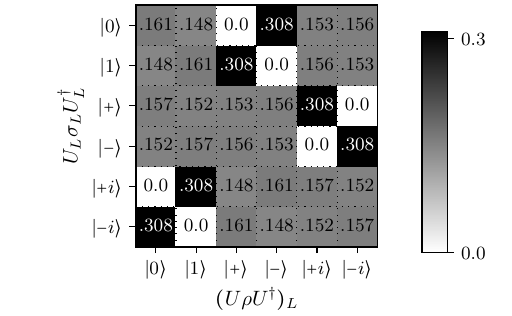}
    }
    \caption{\label{fig:matrix_1q}Geometric interpretation of loss function variants for single-qubit operations. Subplot (a) shows the pairwise fidelities between target 2-design states, i.e., the first summand of \cref{eq:operation_error_extended}. This also nicely visualizes the three variants: $\mathcal{O}_{\mathrm{diag}}$ only considers the diagonal elements; $\mathcal{O}_{\mathrm{block}}$ additionally considers mutually orthogonal states, i.e., a block-diagonal structure; $\mathcal{O}_{\mathrm{full}}$ incorporates the full Gram matrix. In (b), we visualize the prediction values for a random initialization when training a transversal realization of the logical $S$ on the $[[7,1,3]]$ code. While the block structure is also prevalent, ordering is scrambled, and overall values are lower.}
\end{figure}

\begin{figure}[tb]
    \centering
    \includegraphics{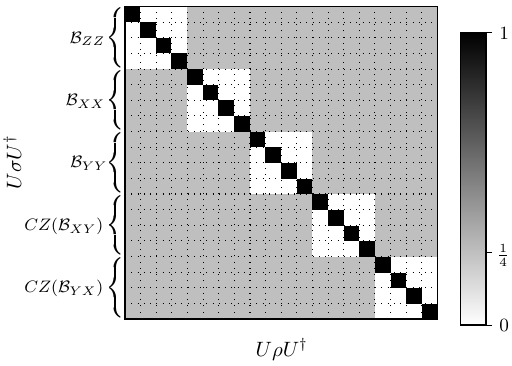}
    \caption{\label{fig:matrix_2q}Geometric interpretation of loss function variants for two-qubit operations. It shows the pairwise fidelities between target 2-design states, i.e., the first summand of \cref{eq:operation_error_extended}. 
    All loss variants are defined equivalently to the single-qubit version in \cref{fig:matrix_1q}, with block-size $4 \times 4$ for $\mathcal{O}_{\mathrm{block}}$.}
\end{figure}

In this appendix, we give a geometric interpretation of the different loss variants introduced in~\cref{subsec:method_loss} and relate them to the structure of 2-designs for single- and two-qubit logical operations. For single-qubit logical gates $U\in\mathrm{U}(2)$ we recall the logical 2-design
\begin{align}
  \mathcal{S}^{(1)}
  =
  \bigl\{ \ket{0}, \ket{1}, \ket{+}, \ket{-}, \ket{+i}, \ket{-i} \bigr\}.
\end{align}
On the Bloch sphere, these six pure states form the vertices of a regular unit-size octahedron (compare \cref{subfig:operation_loss_baseline}). Under the target logical operation $U$, this octahedron is rigidly rotated or reflected, defining the \emph{target} images of the 2-design states, shown as gray octahedrons in Fig.~\ref{fig:operation_loss_training}. Given a parameterized physical realization $U_L(\Psi)$, the corresponding encoded outputs of the 2-design form an interlaced, not necessarily regular, octahedron (black), which represents the \emph{prediction} of the current ansatz. \Cref{fig:operation_loss_training} illustrates this picture for training a logical $S$ gate on the $[[7,1,3]]$ Steane code: from epoch to epoch, the black octahedron is expanded until it coincides with the gray one, i.e., until $U_L(\Psi)$ reproduces the action of $U$ on all 2-design states.

To formalize this, \cref{subsec:method_loss} introduced pairwise operation errors $\Delta_{F}(\rho,\sigma;C;\Psi)$ between logical pure states $\rho,\sigma$ and three natural choices of pair sets,
\begin{align}
  \mathcal{P}_{\mathrm{diag}}
  &= \bigl\{ (\rho,\rho) \mid \rho\in\mathcal{S} \bigr\},\\
  \mathcal{P}_{\mathrm{block}}
  &= \mathcal{P}_{\mathrm{diag}}
     \cup \bigl\{ (\rho,\sigma) \mid \rho\perp\sigma;\; \rho,\sigma\in\mathcal{S} \bigr\},\\
  \mathcal{P}_{\mathrm{full}}
  &= \bigl\{ (\rho,\sigma) \mid \rho,\sigma\in\mathcal{S} \bigr\},
\end{align}
where $\mathcal{S}$ denotes the underlying 2-design on the logical subsystem.
The associated average-case losses $\mathcal{O}_{\mathrm{diag}}$, $\mathcal{O}_{\mathrm{block}}$, and $\mathcal{O}_{\mathrm{full}}$ thus differ only in which state pairs $\rho,\sigma$ they constrain.
A convenient way to visualize these choices is via Gram matrices of pairwise fidelities.
For a fixed target logical gate $U$ and a 2-design $\mathcal{S}=\{\rho_1,\dots,\rho_m\}$ we define the \emph{target} Gram matrix $G^{\mathrm{target}}$ with entries
\begin{equation}
  G^{\mathrm{target}}_{ij}
  = F\bigl(U\rho_i U^\dagger,\, U\rho_j U^\dagger\bigr),
\end{equation}
and, for a given ansatz $U_L(\Psi)$, the \emph{prediction} Gram matrix $G^{\mathrm{pred}}$ with entries
\begin{equation}
  G^{\mathrm{pred}}_{ij}
  = F\bigl((U\rho_i U^\dagger)_L,\,
           U_L(\Psi)\rho_{j,L} U_L(\Psi)^\dagger\bigr),
\end{equation}
where $\rho_{j,L}$ and $(U\rho_i U^\dagger)_L$ are the encoded logical states, compare \cref{subsec:method_single_qubit}.

In \cref{fig:matrix_1q}, we show these matrices for the single-qubit setup, where the target Gram matrix has a particularly simple structure: all diagonal entries are $1$, mutually orthogonal states yield $0$, and all remaining entries are $1/2$. Geometrically, this structure encodes that the six Bloch vectors form a regular octahedron with three pairwise orthogonal pairs. The three loss variants can then be understood as selecting different subsets of matrix entries: $\mathcal{O}_{\mathrm{diag}}$ uses only the diagonal entries $G_{ii}$, $\mathcal{O}_{\mathrm{block}}$ augments this by all entries corresponding to orthogonal pairs $(\rho,\sigma)$, i.e., the $2\times2$ blocks along the diagonal, and $\mathcal{O}_{\mathrm{full}}$ uses all entries of the Gram matrix. The implications of these choices for the optimization landscape and local minima are examined in more detail in \cref{app:loss}.

In \cref{subfig:matrix_steane_S_1q}, we show a typical prediction Gram matrix $G^{\mathrm{pred}}$ at a random initialization before training a logical $S$ gate on the $[[7,1,3]]$ code. Its entries are generally smaller in magnitude and lack the clear pattern of $G^{\mathrm{target}}$, but an approximate block structure can already be discerned, albeit permuted. The training process can be viewed as aligning $G^{\mathrm{pred}}$ with $G^{\mathrm{target}}$ on the subset of entries selected by $\mathcal{P}$. Since $G^{\mathrm{target}}$ depends only on the logical 2-design and the chosen gate $U$, it is static and can be pre-computed once. In contrast, $G^{\mathrm{pred}}$ must be recomputed in every optimization step. On the single-qubit 2-design, additional symmetries in the $2\times 2$ sub-blocks corresponding to the pairs $\{\ket{0},\ket{1}\}$, $\{\ket{+},\ket{-}\}$, and $\{\ket{+i},\ket{-i}\}$ can be exploited to reduce this cost.

The same Gram-matrix picture extends directly to two-qubit logical gates $U\in\mathrm{U}(4)$. We use the two-qubit 2-design
\begin{align}
    \begin{split}
        \mathcal{S}^{(2)}
        =
        \bigl\{&
        \ket{00}, \ket{01}, \ket{10}, \ket{11}, \\
        &\ket{++}, \ket{+-}, \ket{-+}, \ket{--}, \\
        &\ket{{+i}{+i}}, \ket{{+i}{-i}}, \ket{{-i}{+i}}, \ket{{-i}{-i}}, \\
        &\tfrac{1}{2}\!\left(\ket{00}+i\ket{01}+\ket{10}-i\ket{11}\right),
        \\
        &\tfrac{1}{2}\!\left(\ket{00}-i\ket{01}+\ket{10}+i\ket{11}\right),
        \\
        &\tfrac{1}{2}\!\left(\ket{00}+i\ket{01}-\ket{10}+i\ket{11}\right),
        \\
        &\tfrac{1}{2}\!\left(\ket{00}-i\ket{01}-\ket{10}-i\ket{11}\right),
        \\
        &\tfrac{1}{2}\!\left(\ket{00}+\ket{01}+i\ket{10}-i\ket{11}\right),
        \\
        &\tfrac{1}{2}\!\left(\ket{00}-\ket{01}+i\ket{10}+i\ket{11}\right),
        \\
        &\tfrac{1}{2}\!\left(\ket{00}+\ket{01}-i\ket{10}+i\ket{11}\right),
        \\
        &\tfrac{1}{2}\!\left(\ket{00}-\ket{01}-i\ket{10}-i\ket{11}\right)
        \bigr\}.
    \end{split}
\end{align}
The corresponding target Gram matrix is shown in \cref{fig:matrix_2q}. Its structure consists of five natural $4\times 4$ blocks associated with the computational basis, the product $X$ basis, the product $Y$ basis, and the two entangled bases obtained from $CZ(\mathcal{B}_{XY})$ and $CZ(\mathcal{B}_{YX})$. We keep the definitions of $\mathcal{O}_{\mathrm{diag}}$, $\mathcal{O}_{\mathrm{block}}$, and $\mathcal{O}_{\mathrm{full}}$ unchanged and interpret the block-diagonal loss as constraining the five $4\times 4$ diagonal blocks of Fig.~\ref{fig:matrix_2q}. In this setting,
$\mathcal{O}_{\mathrm{block}}$ enforces normalization and orthogonality within each basis, while leaving the mutual-unbiased overlaps between different bases unconstrained. The numerical success statistics for all three variants are reported in \cref{app:empirical}.


\section{\label{app:loss}Training Dynamics and Loss Landscape}

\begin{figure}[tb]
    \centering
    \subfigure[\label{subfig:operation_curve_diag}Sample training curves with diagonal loss function $\mathcal{O}_{\mathrm{diag}}$.]{
        \includegraphics{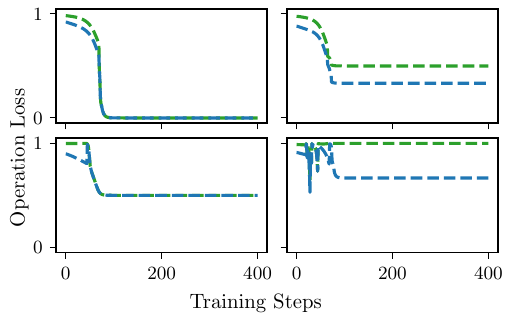}
    }\\
        \subfigure[\label{subfig:operation_curve_block}Sample training curves with block-diagonal loss function $\mathcal{O}_{\mathrm{block}}$.]{
        \includegraphics{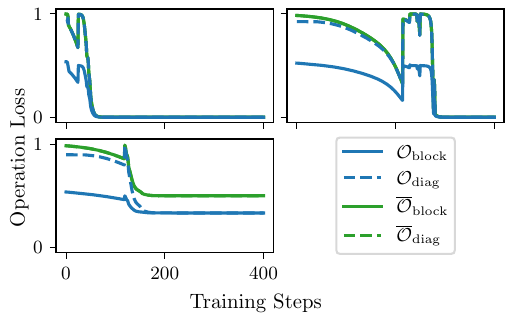}
    }
    \caption{\label{fig:operation_curve}Training dynamics for different random initializations of a transversal $S$ operation on the $[[7,1,3]]$ code. In {a}, we visualize four samples of using the original $\mathcal{O}_{\mathrm{diag}}$ loss, and additionally show the worst-case counterpart $\overline{\mathcal{O}}_{\mathrm{diag}}$. A large fraction of runs converge to local optima; details see \cref{tab:operation_loss}. The ordering of the subplots here corresponds to these local optima visualized on the 2-design octahedron in \cref{fig:operation_loss_subopt}. In (b), we train with the modified $\mathcal{O}_{\mathrm{block}}$ loss, but additionally visualize the other variant. Interestingly, $\overline{\mathcal{O}}_{\mathrm{diag}} = \overline{\mathcal{O}}_{\mathrm{block}}$ holds for almost all observed training instances, and also $\mathcal{O}_{\mathrm{block}}$ converges with $\mathcal{O}_{\mathrm{diag}}$. Therefore, we conclude that the modified loss function produces the desired results while avoiding loss plateaus during training.}
\end{figure}

\begin{figure}[tb]
    \centering
    \includegraphics{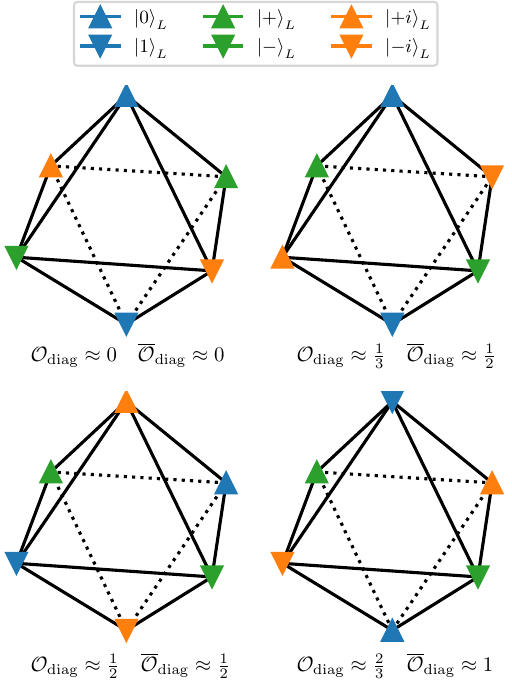}
    \caption{\label{fig:operation_loss_subopt}Illustration of convergence to local minima on the 2-design octahedron. We use the $\mathcal{O}_{\mathrm{diag}}$ loss to train a transversal realization of a logical $S$ operation on the $[[7,1,3]]$ code. A large fraction of runs converge to local optima; details see \cref{tab:operation_loss}. While the predictions span the full 2-design octahedron, the assignment of actual 2-design states is wrong. The order of the four samples corresponds to the training curves shown in \cref{subfig:operation_curve_diag}. As argued in this section and \cref{app:loss}, the block-diagonal version $\mathcal{O}_{\mathrm{block}}$ breaks these degeneracies and drastically improves the convergence rate.}
\end{figure}

We now illustrate the typical training dynamics for the different loss variants and analyze the structure of local minima in more detail. Throughout this appendix, we use the logical $S$ gate on the $[[7,1,3]]$ code as a representative example, with a transversal ansatz (single repetition) and an L-BFGS optimizer with $20$ steps per epoch.

In \cref{fig:operation_curve}, we show representative training curves for several random initializations of the transversal parameters. In \cref{subfig:operation_curve_diag}, we optimize using the diagonal loss $\mathcal{O}_{\mathrm{diag}}$, but also track the corresponding worst-case version $\overline{\mathcal{O}}_{\mathrm{diag}}$. Only a minority of runs converge below the target threshold of $10^{-5}$, see \cref{app:empirical} for absolute numbers. Most trajectories saturate at plateaus corresponding to distinct local optima, with three typical value combinations:
\begin{align}
    \mathcal{O}_{\mathrm{diag}}\approx \tfrac{1}{3},&\quad \overline{\mathcal{O}}_{\mathrm{diag}}\approx \tfrac{1}{2} \\
    \mathcal{O}_{\mathrm{diag}}\approx \tfrac{1}{2},&\quad \overline{\mathcal{O}}_{\mathrm{diag}}\approx \tfrac{1}{2} \\
    \mathcal{O}_{\mathrm{diag}}\approx \tfrac{2}{3},&\quad \overline{\mathcal{O}}_{\mathrm{diag}}\approx 1,
\end{align}
The geometric nature of these local minima becomes clear in \cref{fig:operation_loss_subopt}, where the corresponding final predictions are plotted on the 2-design octahedron. In all cases, the learned predictions span the full space and differ from the target only by permutations and sign flips of the 2-design vertices. In other words, the ansatz has learned the correct orbit of the logical $S$ gate on the Bloch sphere, but has assigned the individual 2-design states to wrong positions. Because $\mathcal{O}_{\mathrm{diag}}$ only compares each state to its own target image, these mis-labeled but geometrically correct configurations can form genuine local minima: they incur an intermediate and approximately uniform error on several 2-design states, which the optimizer cannot easily escape.

\begin{table}[tb]
    \caption{\label{tab:operation_loss}Full numerical statistics for discovering transversal logical operations of stabilizer codes underlying \cref{subfig:operation_loss}. We compare $22$ one-qubit and $7$ two-qubit setups, using the three loss variants $\mathcal{O}_{\mathrm{diag}},\mathcal{O}_{\mathrm{block}},\mathcal{O}_{\mathrm{full}}$ as defined in \cref{subsec:method_loss}. We train with $100$ random initializations for a transversal ansatz (no repetitions) each, and report the number of runs that converged to a worst-case loss $\overline{\mathcal{O}}_{\mathcal{S}} \leq 10^{-5}$. We use $-$ to mark combinations where no transversal realization of the respective logical operation exists.}
    \begin{ruledtabular}
        \begin{tabular}{cccc|ccccccccc}
            & \multirow{2}{*}{\textbf{Code}} & \multirow{2}{*}{\textbf{Loss}} && \multicolumn{8}{c}{\textbf{Transversal Operation}} \\
            &&&& \textbf{X} & \textbf{Z} & \textbf{H} & \textbf{S} & \textbf{T} & \textbf{CX} & \textbf{CZ} & \textbf{CS} \\
            \hline
            &\multirow{4}{*}{[[$3,1,1$]]} & $\mathcal{O}_{\mathrm{diag}}$ && $2$ & $100$ & $-$ & $95$ & $91$ & $1$ & $100$ & $100$ \\
            && $\mathcal{O}_{\mathrm{block}}$ && $69$ & $100$ & $-$ & $100$ & $100$ & $100$ & $100$ & $100$ \\
            && $\mathcal{O}_{\mathrm{full}}$ && $28$ & $72$ & $-$ & $53$ & $41$ & $17$ & $68$ & $44$ \\
            \hline
            &\multirow{4}{*}{[[$4,1,2$]]} & $\mathcal{O}_{\mathrm{diag}}$ && $18$ & $77$ & $-$ & $-$ & $-$ & $7$ & $-$ & $-$ \\
            && $\mathcal{O}_{\mathrm{block}}$ && $96$ & $97$ & $-$ & $-$ & $-$ & $99$ & $-$ & $-$ \\
            && $\mathcal{O}_{\mathrm{full}}$ && $15$ & $53$ & $-$ & $-$ & $-$ & $43$ & $-$ & $-$ \\
            \hline
            &\multirow{3}{*}{[[$5,1,3$]]} & $\mathcal{O}_{\mathrm{diag}}$ && $34$ & $46$ & $-$ & $-$ & $-$ & $-$ & $-$ & $-$ \\
            && $\mathcal{O}_{\mathrm{block}}$ && $100$ & $100$ & $-$ & $-$ & $-$ & $-$ & $-$ & $-$ \\
            && $\mathcal{O}_{\mathrm{full}}$ && $27$ & $29$ & $-$ & $-$ & $-$ & $-$ & $-$ & $-$ \\
            \hline
            &\multirow{4}{*}{[[$7,1,3$]]} & $\mathcal{O}_{\mathrm{diag}}$ && $12$ & $45$ & $2$ & $26$ & $-$ & $8$ & $43$ & $-$ \\
            && $\mathcal{O}_{\mathrm{block}}$ && $93$ & $89$ & $21$ & $92$ & $-$ & $47$ & $57$ & $-$ \\
            && $\mathcal{O}_{\mathrm{full}}$ && $6$ & $19$ & $2$ & $10$ & $-$ & $15$ & $11$ & $-$ \\
            \hline
            &\multirow{4}{*}{[[$9,1,3$]]} & $\mathcal{O}_{\mathrm{diag}}$ && $79$ & $8$ & $-$ & $-$ & $-$ & $~8$\footnotemark[1] & $-$ & $-$ \\
            && $\mathcal{O}_{\mathrm{block}}$ && $96$ & $53$ & $-$ & $-$ & $-$ & $~92$\footnotemark[1] & $-$ & $-$ \\
            && $\mathcal{O}_{\mathrm{full}}$ && $70$ & $8$ & $-$ & $-$ & $-$ & $~9$\footnotemark[1] & $-$ & $-$ \\
            \hline
            &\multirow{3}{*}{[[$10,1,4$]]} & $\mathcal{O}_{\mathrm{diag}}$ && $55$ & $60$ & $-$ & $-$ & $-$ & $-$ & $-$ & $-$ \\
            && $\mathcal{O}_{\mathrm{block}}$ && $98$ & $99$ & $-$ & $-$ & $-$ & $-$ & $-$ & $-$ \\
            && $\mathcal{O}_{\mathrm{full}}$ && $42$ & $40$ & $-$ & $-$ & $-$ & $-$ & $-$ & $-$ \\
            \hline
            &\multirow{3}{*}{[[$11,1,5$]]} & $\mathcal{O}_{\mathrm{diag}}$ && $52$ & $48$ & $-$ & $-$ & $-$ & $-$ & $-$ & $-$ \\
            && $\mathcal{O}_{\mathrm{block}}$ && $94$ & $89$ & $-$ & $-$ & $-$ & $-$ & $-$ & $-$ \\
            && $\mathcal{O}_{\mathrm{full}}$ && $34$ & $37$ & $-$ & $-$ & $-$ & $-$ & $-$ & $-$ \\
            \hline
            &\multirow{4}{*}{[[$15,1,3$]]} & $\mathcal{O}_{\mathrm{diag}}$ && $2$ & $8$ & $-$ & $14$ & $16$ & $~-$\footnotemark[2] & $~-$\footnotemark[2] & $~-$\footnotemark[2] \\
            && $\mathcal{O}_{\mathrm{block}}$ && $1$ & $41$ & $-$ & $41$ & $41$ & $~-$\footnotemark[2] & $~-$\footnotemark[2] & $~-$\footnotemark[2] \\
            && $\mathcal{O}_{\mathrm{full}}$ && $3$ & $9$ & $-$ & $7$ & $7$ & $~-$\footnotemark[2] & $~-$\footnotemark[2] & $~-$\footnotemark[2] \\
        \end{tabular}
    \end{ruledtabular}
    \footnotetext[1]{Control and target order inverted for physical realization.}
    \footnotetext[2]{Existence of transversal realization is known, but excluded from evaluation due to computational complexity.}
\end{table}

\begin{table}[tb]
    \caption{\label{tab:operation_repeat}Full numerical statistics for discovering transversal logical operations of stabilizer codes underlying \cref{subfig:operation_repeat}. We compare $22$ one-qubit and $7$ two-qubit setups, using the loss variant $\mathcal{O}_{\mathrm{block}}$ as defined in \cref{subsec:method_loss}. We train with $100$ random initializations for a transversal ansatz with up to $5$ repetitions each, and report the number of runs that converged to a worst-case loss $\overline{\mathcal{O}}_{\mathcal{S}} \leq 10^{-5}$. We use $-$ to mark combinations where no transversal realization of the respective logical operation exists.}
    \begin{ruledtabular}
        \begin{tabular}{cccc|ccccccccc}
            & \multirow{2}{*}{\textbf{Code}} & \multirow{2}{*}{\textbf{Repeat}} && \multicolumn{8}{c}{\textbf{Transversal Operation}} \\
            &&&& \textbf{X} & \textbf{Z} & \textbf{H} & \textbf{S} & \textbf{T} & \textbf{CX} & \textbf{CZ} & \textbf{CS} \\
            \hline
            &\multirow{5}{*}{[[$3,1,1$]]} & $1$ && $69$ & $100$ & $-$ & $100$ & $100$ & $100$ & $100$ & $100$ \\
            && $2$ && $100$ & $100$ & $-$ & $100$ & $100$ & $100$ & $100$ & $100$ \\
            && $3$ && $100$ & $100$ & $-$ & $100$ & $100$ & $100$ & $100$ & $100$ \\
            && $4$ && $100$ & $100$ & $-$ & $100$ & $100$ & $100$ & $100$ & $100$ \\
            && $5$ && $100$ & $100$ & $-$ & $100$ & $100$ & $100$ & $100$ & $100$ \\
            \hline
            &\multirow{5}{*}{[[$4,1,2$]]} & $1$ && $96$ & $97$ & $-$ & $-$ & $-$ & $99$ & $-$ & $-$ \\
            && $2$ && $99$ & $100$ & $-$ & $-$ & $-$ & $99$ & $-$ & $-$ \\
            && $3$ && $99$ & $100$ & $-$ & $-$ & $-$ & $99$ & $-$ & $-$ \\
            && $4$ && $99$ & $100$ & $-$ & $-$ & $-$ & $100$ & $-$ & $-$ \\
            && $5$ && $99$ & $100$ & $-$ & $-$ & $-$ & $100$ & $-$ & $-$ \\
            \hline
            &\multirow{5}{*}{[[$5,1,3$]]} & $1$ && $100$ & $100$ & $-$ & $-$ & $-$ & $-$ & $-$ & $-$ \\
            && $2$ && $100$ & $100$ & $-$ & $-$ & $-$ & $-$ & $-$ & $-$ \\
            && $3$ && $100$ & $100$ & $-$ & $-$ & $-$ & $-$ & $-$ & $-$ \\
            && $4$ && $100$ & $100$ & $-$ & $-$ & $-$ & $-$ & $-$ & $-$ \\
            && $5$ && $100$ & $100$ & $-$ & $-$ & $-$ & $-$ & $-$ & $-$ \\
            \hline
            &\multirow{5}{*}{[[$7,1,3$]]} & $1$ && $93$ & $89$ & $21$ & $92$ & $-$ & $47$ & $57$ & $-$ \\
            && $2$ && $95$ & $89$ & $87$ & $95$ & $-$ & $85$ & $98$ & $-$ \\
            && $3$ && $92$ & $97$ & $91$ & $92$ & $-$ & $98$ & $99$ & $-$ \\
            && $4$ && $94$ & $93$ & $87$ & $94$ & $-$ & $97$ & $99$ & $-$ \\
            && $5$ && $89$ & $90$ & $92$ & $94$ & $-$ & $99$ & $100$ & $-$ \\
            \hline
            &\multirow{5}{*}{[[$9,1,3$]]} & $1$ && $96$ & $53$ & $-$ & $-$ & $-$ & $~92$\footnotemark[1] & $-$ & $-$ \\
            && $2$ && $98$ & $99$ & $-$ & $-$ & $-$ & $~96$\footnotemark[1] & $-$ & $-$ \\
            && $3$ && $99$ & $99$ & $-$ & $-$ & $-$ & $~97$\footnotemark[1] & $-$ & $-$ \\
            && $4$ && $100$ & $98$ & $-$ & $-$ & $-$ & $~97$\footnotemark[1] & $-$ & $-$ \\
            && $5$ && $100$ & $96$ & $-$ & $-$ & $-$ & $~100$\footnotemark[1] & $-$ & $-$ \\
            \hline
            &\multirow{5}{*}{[[$10,1,4$]]} & $1$ && $98$ & $99$ & $-$ & $-$ & $-$ & $-$ & $-$ & $-$ \\
            && $2$ && $54$ & $48$ & $-$ & $-$ & $-$ & $-$ & $-$ & $-$ \\
            && $3$ && $95$ & $78$ & $-$ & $-$ & $-$ & $-$ & $-$ & $-$ \\
            && $4$ && $43$ & $56$ & $-$ & $-$ & $-$ & $-$ & $-$ & $-$ \\
            && $5$ && $96$ & $99$ & $-$ & $-$ & $-$ & $-$ & $-$ & $-$ \\
            \hline
            &\multirow{5}{*}{[[$11,1,5$]]} & $1$ && $94$ & $89$ & $-$ & $-$ & $-$ & $-$ & $-$ & $-$ \\
            && $2$ && $6$ & $13$ & $-$ & $-$ & $-$ & $-$ & $-$ & $-$ \\
            && $3$ && $28$ & $53$ & $-$ & $-$ & $-$ & $-$ & $-$ & $-$ \\
            && $4$ && $27$ & $21$ & $-$ & $-$ & $-$ & $-$ & $-$ & $-$ \\
            && $5$ && $94$ & $90$ & $-$ & $-$ & $-$ & $-$ & $-$ & $-$ \\
            \hline
            &\multirow{5}{*}{[[$15,1,3$]]} & $1$ && $1$ & $41$ & $-$ & $41$ & $41$ & $~-$\footnotemark[2] & $~-$\footnotemark[2] & $~-$\footnotemark[2] \\
            && $2$ && $9$ & $47$ & $-$ & $41$ & $49$ & $~-$\footnotemark[2] & $~-$\footnotemark[2] & $~-$\footnotemark[2] \\
            && $3$ && $11$ & $25$ & $-$ & $29$ & $15$ & $~-$\footnotemark[2] & $-~$\footnotemark[2] & $~-$\footnotemark[2] \\
            && $4$ && $16$ & $34$ & $-$ & $34$ & $40$ & $~-$\footnotemark[2] & $~-$\footnotemark[2] & $~-$\footnotemark[2] \\
            && $5$ && $4$ & $51$ & $-$ & $51$ & $51$ & $~-$\footnotemark[2] & $~-$\footnotemark[2] & $~-$\footnotemark[2] \\
        \end{tabular}
    \end{ruledtabular}
    \footnotetext[1]{Control and target order inverted for physical realization.}
    \footnotetext[2]{Existence of transversal realization is known, but excluded from evaluation due to computational complexity.}
\end{table}

Augmenting the loss to the block-diagonal variant $\mathcal{O}_{\mathrm{block}}$ breaks most of these degeneracies. By additionally constraining pairs of mutually orthogonal 2-design states, $\mathcal{O}_{\mathrm{block}}$ enforces not only that each state is mapped onto the correct orbit, but also that orthogonality relations between 2-design states are preserved. This rules out permutations that scramble which states are orthogonal to which, and thus eliminates many of the plateau solutions observed with $\mathcal{O}_{\mathrm{diag}}$. In \cref{subfig:operation_curve_block}, we show sample training curves when optimizing using $\mathcal{O}_{\mathrm{block}}$. Convergence to good solutions becomes substantially more likely (compare \cref{app:empirical}), and in almost all successful runs, we empirically find
\begin{align}
    \overline{\mathcal{O}}_{\mathrm{block}} = \overline{\mathcal{O}}_{\mathrm{diag}},
\end{align}
as well as
\begin{align}
  \mathcal{O}_{\mathrm{block}}\approx \mathcal{O}_{\mathrm{diag}},
\end{align}
once the model is near an optimum. In other words, the block-diagonal loss is a faithful proxy for the original diagonal objective in the neighborhood of a good solution, while avoiding many of the sub-optimal plateaus during optimization.

On paper, the full-pair loss $\mathcal{O}_{\mathrm{full}}$ should also be able to remove these degeneracies, since it constrains the complete Gram matrix of pairwise fidelities. In practice, however, we find that training with $\mathcal{O}_{\mathrm{full}}$ is noticeably harder: the optimization landscape becomes more intricate, and success probabilities are similar to those of $\mathcal{O}_{\mathrm{diag}}$ but clearly inferior to $\mathcal{O}_{\mathrm{block}}$, compare \cref{subfig:operation_loss} and \cref{app:empirical}.
Taken together, these observations motivate our choice of $\mathcal{O}_{\mathrm{block}}$ as the default loss formulation used throughout this work.


\section{\label{app:empirical}Full Numerical Success Statistics for Loss Function Variants}

This appendix documents the full numerical success statistics underlying the aggregate results reported in \cref{subsec:empirical_loss}. We first describe the experimental protocol and then summarize the behavior of different loss variants and the effect of overparameterizing transversal ansätze.

We evaluate the learning procedure on the standard $[[n,1,d]]$ stabilizer codes and compiled in \cref{tab:stabilizer_transversal}. For each code, we attempt to learn transversal physical realizations of the intra-block single-logical-qubit gates
\begin{align}
  \{ X_L, Z_L, H_L, S_L, T_L \},
\end{align}
and the inter-block two-logical-qubit gates
\begin{align}
  \{ CX_L, CZ_L, CS_L \},
\end{align}
whenever a strictly-transversal or transversal implementation is known to exist. In total, this yields $22$ single-qubit and $10$ two-qubit code–operation pairs. Three of the two-qubit instances on the $[[15,1,3]]$ Reed–Muller code would require simulating $30$ physical qubits and are therefore omitted due to the limitations of our current state-vector-based pipeline, leaving $7$ two-qubit instances in the numerical sweeps. Training was conducted using an L-BFGS optimizer with $20$ internal steps per epoch and $50$ epochs in total. For each configuration (code, gate, loss variant), we perform $100$ runs with independent random initializations of the transversal parameters. As a quality criterion, we use the worst-case 2-design loss, as also throughout the main part of this paper, and declare a run successful if
\begin{align}
  \overline{\mathcal{O}}_{\mathcal{S}} \le 10^{-5}.
\end{align}

In \cref{tab:operation_loss}, we report, for each code–gate pair, the number of successful runs when training with the losses $\mathcal{O}_{\mathrm{diag}}$, $\mathcal{O}_{\mathrm{block}}$, and $\mathcal{O}_{\mathrm{full}}$. These data underlie the histograms in \cref{subfig:operation_loss}. Two key empirical observations emerge: first, over all code–operation combinations, the block-diagonal loss $\mathcal{O}_{\mathrm{block}}$ consistently achieves the highest success percentages. For small codes up to five qubits, success rates are typically close to $100\%$, and also for larger codes, $\mathcal{O}_{\mathrm{block}}$ dominates the other variants in median and average performance. Second, across all instances, the framework never identifies a transversal realization when the ground truth (see \cref{tab:stabilizer_transversal}) states that none exists. This absence of false positives further validates the correctness of the loss formulation and optimization pipeline.

To investigate the effect of overparameterizing the transversal ansatz without changing its logical depth, we study the repeated-layer construction of \cref{eq:transversal_1q_over}. Concretely, we repeat the transversal layer of single-qubit (or controlled-single-qubit) gates $r\in\{1,\dots,5\}$ times with independent parameters. After training, these repetitions can be analytically compiled back into a single gate per physical qubit, so the final logical operation remains depth-one. In \cref{tab:operation_repeat}, we report success counts for all code–operation pairs when training with the block-diagonal loss $\mathcal{O}_{\mathrm{block}}$ and respective ansatz repetition rates. The corresponding aggregated statistics underlie \cref{subfig:operation_repeat}. For almost all codes and gates, increasing $r$ substantially improves success rates for both single- and two-qubit operations, with performance saturating around $r=3$ to $r=5$. For all codes up to $n=11$ physical qubits, the success probabilities with moderate repetition exceed $90\%$. For the $[[15,1,3]]$ code, success rates drop to roughly $50\%$ for most gates, and in the special case of the logical $X_L$ gate, even lower rates are observed. This indicates potential scalability challenges of the current state-vector-based implementation when moving to larger code sizes, and suggests that more advanced simulation techniques, such as tensor-network-based methods for topological codes (e.g., PEPO representations~\cite{darmawan2017tensor,huang2025robust}), will be required in future work.

\bigskip
\noindent
Combining the geometric picture of App.~\ref{app:geometric} with the empirical evidence in \cref{tab:operation_loss,tab:operation_repeat}, we conclude that the block-diagonal loss $\mathcal{O}_{\mathrm{block}}$ offers the best trade-off between optimization stability and computational cost. It faithfully approximates the original diagonal loss near good solutions, while avoiding the most problematic local minima. We therefore adopt $\mathcal{O}_{\mathrm{block}}$ as the standard loss formulation in this work and use it throughout \cref{sec:vareftqc} for the \gls{vareftqc} co-design algorithm.

\bibliography{paper}

\end{document}